\documentclass[10pt,twocolumn]{article}
\usepackage{graphicx} 
\usepackage{cite}
\usepackage{fontawesome}
\usepackage{amsmath,amsfonts,amscd}
\usepackage{amssymb}
\usepackage{fancyhdr}
\usepackage{mathrsfs}
\usepackage{textcomp}
\usepackage{multirow}
\usepackage{amsthm}
\usepackage[T1]{fontenc}
\usepackage{enumitem}
\usepackage{algorithm,algorithmicx,algpseudocode}
\usepackage{listings}
\usepackage{booktabs}
\usepackage{threeparttable}
\usepackage{color}
\usepackage{titlesec}
\usepackage{subcaption}
\usepackage{svg}
\usepackage{pifont}
\usepackage{hyperref} 
\usepackage{setspace}
\usepackage{xcolor}

\hypersetup{
    colorlinks=true,         
    linkcolor=red,         
    citecolor={green!70!black},         
    urlcolor=black,          
    filecolor=black,         
    pdfborder={0 0 0}        
}

\title{\LARGE\bfseries\sffamily SparkAttention: High-Performance Multi-Head Attention for Large Models on Volta GPU Architecture}
\author{Youxuan Xu, Tong Wu, Shigang Li \thanks{Corresponding Author: shigangli.cs@gmail.com \\ Published in CCF Transactions on High Performance Computing (CCF THPC), 2025 \\ DOI: \url{https://doi.org/10.1007/s42514-024-00211-0} }\ \ , Xueying Wang, Jingjing Wang\\ School of Computer Science, Beijing University of Posts and Telecommunications}

\date{}

\begin{document}

\maketitle

\singlespacing

\thispagestyle{empty}

\begin{abstract} \label{sec:abstract}
Transformer are widely used in various fields such as natural language processing and computer vision. 
However, the training time for large Transformer models can be challenging due to the Multi-Head Attention (MHA) mechanism. 
Especially as models become larger, training becomes more costly.
So it is crucial to utilize various resources for efficient model training. 
Currently, NVIDIA Volta GPU is still widely used. 
However, because the computational shapes supported by Tensor Core Units (TCU) of Volta GPU differ from other GPU architectures, most efforts have not focused on using them to accelerate Transformer training. 
To address this issue, we propose SparkAttention, an acceleration library designed to speed up MHA training on the Volta GPU. 
SparkAttention leverages TCU and kernel fusion to reduce the number of high bandwidth memory (HBM) accesses and overhead.
Our End-to-End experimental results on an NVIDIA V100 GPU show that SparkAttention achieves on average 1.80$\times$ (up to 2.46$\times$) speedup compared to using PyTorch.

\end{abstract}

\section{Introduction} \label{sec:introduction}
In recent years, Transformer~\cite{vaswani2017attention}models have been increasingly applied in the fields of natural language processing\cite{wolf2020transformers,kalyan2021ammus,wolf2019huggingface}, computer vision\cite{wu2020visual,bi2021transformer,liu2021swin}, speech recognition\cite{dong2018speech,gulati2020conformer,zhang2020transformer} and so on. 
These applications have achieved satisfactory success, but this comes at the cost of utilizing a vast number of model parameters.
The growth of model parameters means longer training times and increased energy consumption and carbon emissions\cite{powerai,strubell2020energy,patterson2021carbon}. 
For example, The success of GPT-4\cite{openai2023gpt4} is largely due to its extensive model parameters. However, this also makes the model computationally intensive, requiring significant time and resources to train.
Therefore, it is important to utilize any available resources (Including the Volta\cite{volta} GPU, which is still widely used by many companies) to reduce training time.

\setlength{\tabcolsep}{3pt}
\renewcommand{\arraystretch}{1.25}
\begin{table}[]
\centering
\caption{Supported MMA shapes by FlashAttention-2 and SparkAttention.}
\begin{tabular}{|c|cc|c|}
\hline
\multirow{2}{*}{MMA Shapes}                                & \multicolumn{2}{c|}{GPU Architectures}                                   & \multirow{2}{*}{Library}                                        \\ \cline{2-3}
                                                           & \multicolumn{1}{l|}{Volta} & \multicolumn{1}{l|}{Ampere, Hopper} &                                                                 \\ \hline
m8n8k4                                                     & \multicolumn{1}{c|}{\checkmark}    &         Not optimized                       & \begin{tabular}[c]{@{}c@{}}SparkAttention\\ (ours)\end{tabular} \\ \hline
\begin{tabular}[c]{@{}c@{}}m16n8k8\\ m16n8k16\end{tabular} & \multicolumn{1}{c|}{\ding{55}}     & \checkmark                              & FlashAttention-2                                                \\ \hline
\end{tabular}
\label{tab:mmashapes}
\end{table}
In this paper, we define MHA to include both MHA-Forward and MHA-Backward.
Developing an efficient MHA training library faces three challenges on Volta GPU: \textbf{First}, scalar computations such as softmax become a performance bottleneck for the MHA-Forward in the process of training. 
Because TCU can provide fast matrix computations, while scalar computations can only be performed by CUDA Cores. 
Therefore, increasing the effective time for matrix computations is crucial (this is equivalent to reducing the I/O frequency).
\textbf{Second}, the computation of MHA needs to minimize HBM overhead as much as possible in order to train larger models.
\textbf{Third}, Volta GPU has specific matrix computation shape. 
So, it is necessary to consider how to adapt optimization methods to the Volta GPU.

FlashAttention (FA)\cite{dao2022flashattention} is an algorithm library primarily used to accelerate Transformer training. 
FA utilizes TCU\cite{tensor} to parallelize computations during training based on batch size and multi-head number.
Also, FA takes into account the I/O load during the training process and continuously computes the MHA using techniques like online softmax and two-stage matrix multiplication fusion.
However, FA does not consider the parallel division of sequence lengths for each batch during the computation process.
So, Dao et al. proposed FlashAttention-2 (FA2)\cite{dao2023flashattention} based on FA. 
FA2 significantly improves the utilization of TCU, making it the mainstream algorithm for training Transformer models.
The algorithm proposed by FA2 has been implemented in most Transformer training frameworks.
However, FA2 only supports GPU architectures from Turing onwards, primarily due to differences in Matrix Multiply-Accumulate (MMA)~\cite{mma} shapes as shown in Table \ref{tab:mmashapes}.

Drawing on the algorithmic ideas of FA2, SparkAttention addresses the aforementioned three challenges to some extent.
Specifically, through the customized data exchange strategy that enables online softmax~\cite{milakov2018online,rabe2021self,dao2022flashattention,dao2023flashattention,kitaev2020reformer} and the two-stage matrix multiplication fusion achieved by adjusting the matrix data layout, we have addressed the first and second challenges.
Also, by reconsidering the hardware characteristic of Volta GPU, we propose a high-performance MHA training library to address the third challenge.

Our main contributions are:
\begin{itemize} \small
\item We implement multi-head attention on the Volta GPU architecture with Tensor Core hardware, including support for the MMA instructions with specific shapes.
\item We propose a data exchange strategy to support the online softmax and reduce the data movement between HBM and SRAM.
\item We utilize warp-level layout transform to achieve two-stage matrix multiplication fusion and mitigate the HBM overhead in Transformer model training.
\end{itemize}

Compared to PyTorch\cite{pytorch} (cuBLAS\cite{cublas}), we achieve an average speedup of 4.55$\times$ (up to 9.17$\times$) for MHA-Forward and 3.44$\times$ (up to 7.91$\times$) for MHA-Backward. 
Our End-to-End implementation achieves an average speedup of 1.80$\times$ (up to 2.46$\times$).

\begin{figure}[htbp] \small
  \centering
  \includegraphics[width=\columnwidth]{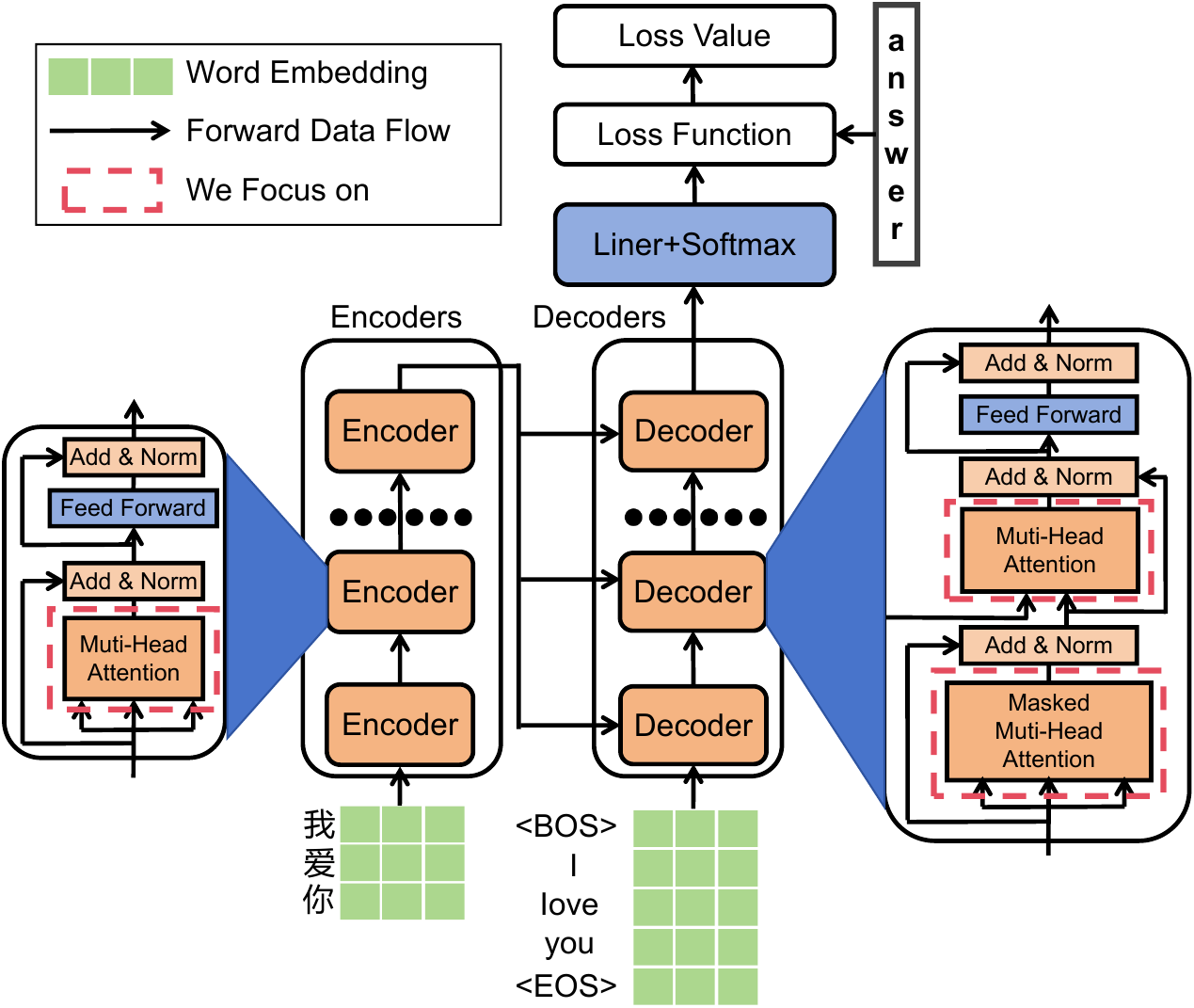}
  \caption{Data flow in the forward computation of a traditional Transformer model training for translation.}
  \label{fig:transformer_example}
\end{figure}

\begin{figure*}[htbp] \small
  \centering
    \begin{subfigure}[b]{0.45\textwidth}
        \includegraphics[width=\textwidth]{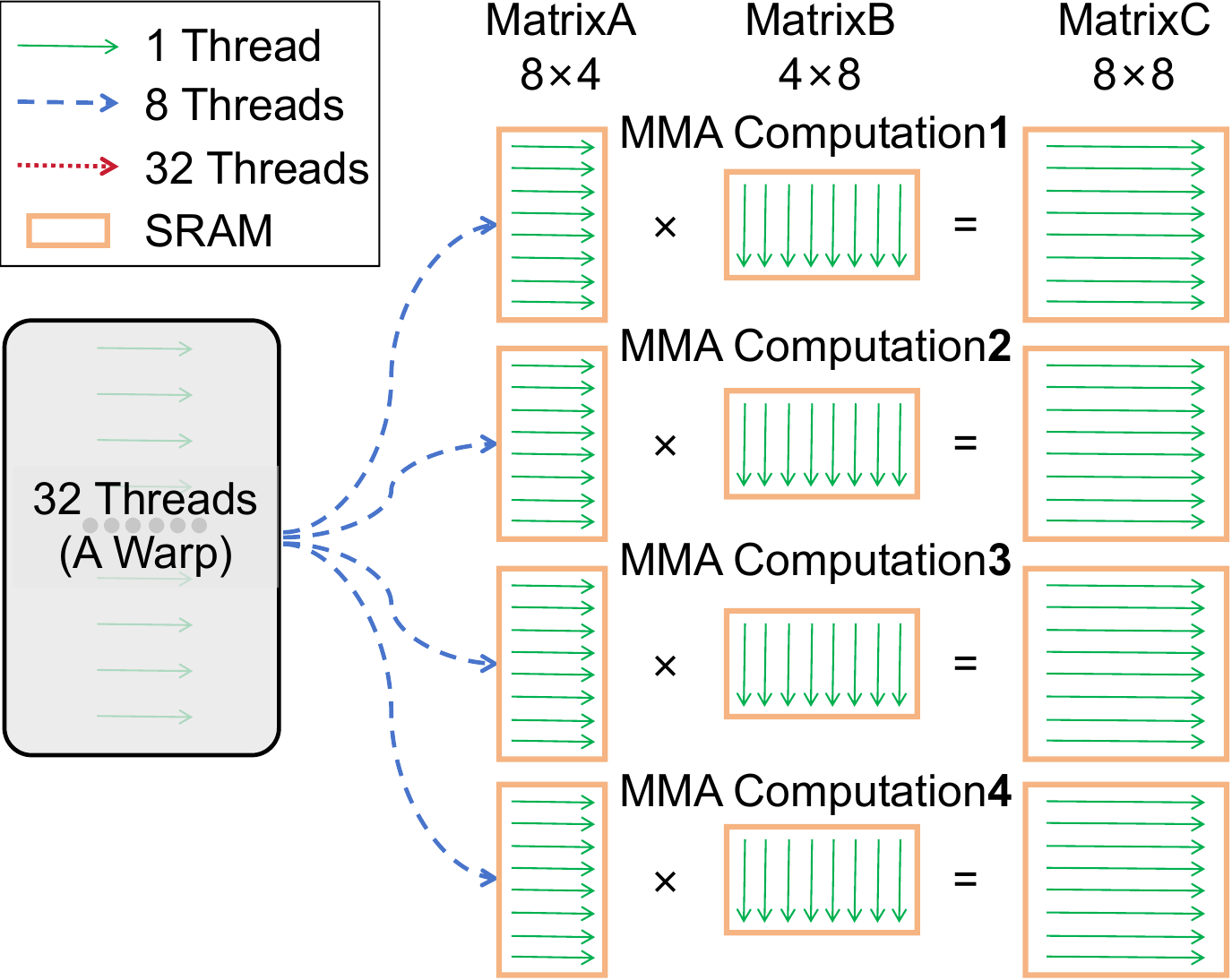}
        \caption{Volta architecture MMA with \textit{m8n8k4}.}
        \label{fig:mma_m8n8k4-a}
    \end{subfigure}
    \hspace{0.0cm}
    \begin{subfigure}[b]{0.45\textwidth}
        \includegraphics[width=\textwidth]{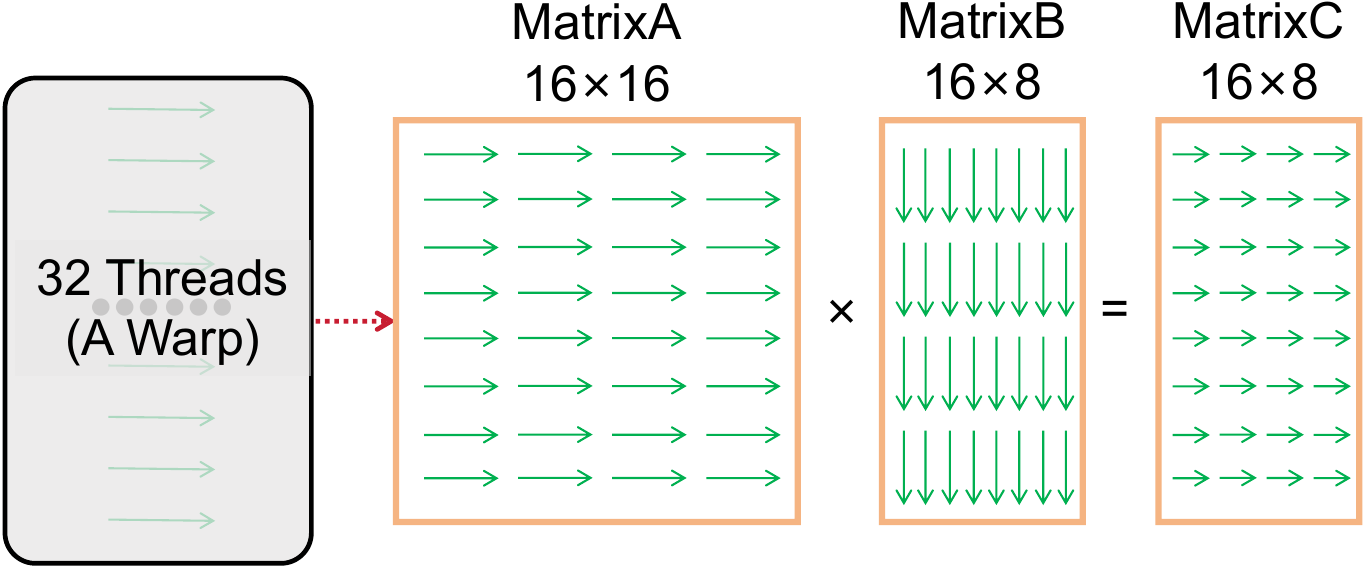}
        \vspace{8mm}
        \caption{Other architectures MMA with \textit{m16n8k16}.}
        \label{fig:mma_m8n8k4-b}
    \end{subfigure}
  \caption{MMA computation of a warp on different GPU architectures.}
  \label{fig:mma_m8n8k4}
\end{figure*}

\section{Background and Motivation} \label{sec:BandM}
\subsection{Transformer Training}
Transformer is widely popular in the field of natural language processing due to its MHA mechanism. 
Therefore, accelerating model training has become a hot topic. 
As shown in Figure \ref{fig:transformer_example}, it describes the forward data flow of the traditional Transformer model.
Specifically, in the Encoder phase, the input consists of the questions from the training set, which are the content that the user needs to translate.
After the computation of each Encoder layer, the final output will be retained for cross-attention computation with the Decoder phase, meaning that the purpose of the Encoder phase is to extract feature information from the input content.
In the Decoder phase, the input consists of the translated results (including the start and end tokens).
After the computation of each Decoder layer, the final output of the Decoder phase is processed through linear layers and softmax to obtain the probability distribution for predicting the next token under the current model parameters. Finally, the forward computation is completed by calculating the loss value.
It is important to note that the Decoder requires two MHA computations. 
The first MHA is a masked computation, meaning that the current token only attends to the tokens that precede it in the Decoder input sequence. 
The second MHA performs cross-attention computation with the final output of the Encoder phase.
Data flow in backward is the reverse of forward. The gradients obtained through backward computation will update all trainable parameters in the model.

The calculation of MHA-Forward is shown in Equation \ref{equ:Transformer}. 
$QKV \in \mathbb{R}^{N \times d}$, where \textit{N} is the length of the input sequence, and \textit{d} is the length of the embedding vector. 
We need to compute the attention output $O \in \mathbb{R}^{N \times d}$. 
For MHA-Forward mechanism, it means dividing the embedding vector into equal parts based on the number of heads (each head generally has its own trainable parameters), computing each head according to Equation \ref{equ:Transformer}, and finally concatenating the results of each head and passing them through a linear layer to obtain the attention output $O$.
\begin{equation} \small \label{equ:Transformer}
\begin{gathered}
    S = QK^T \in \mathbb{R}^{N \times N}\\
    P = softmax(S)/\sqrt{d} \in \mathbb{R}^{N \times N}\\
    O = PV \in \mathbb{R}^{N \times d}\\
\end{gathered}
\end{equation}

\subsection{NVIDIA V100 GPU}

We will introduce the architectural details of the NVIDIA V100 and the specifics of its matrix multiplication calculations using TCU. 
SparkAttention can fully leverage its advantages on GPU with the Volta architecture.

\begin{figure}[htbp] \small
  \centering
  \includegraphics[width=\columnwidth]{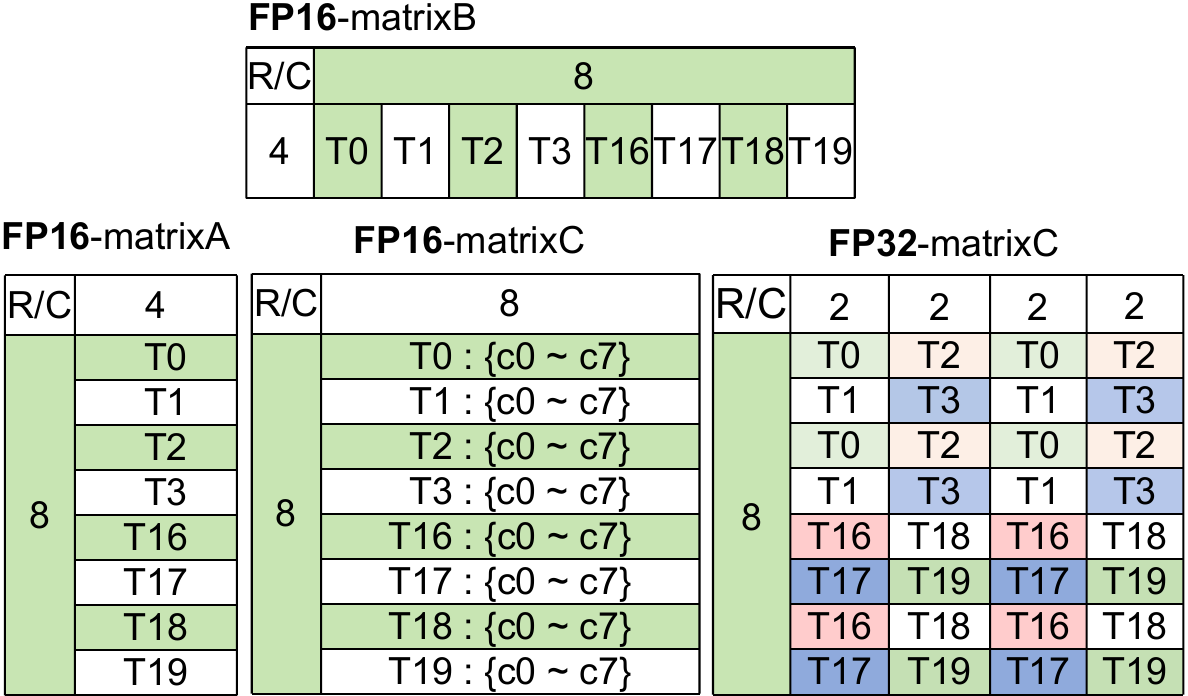}
  \caption{MMA (\textit{m8n8k4}) Computation 1 in Volta architecture.}
  \label{fig:mma_computation1}
\end{figure}
\begin{figure*}[htbp] \small
    \centering
    \begin{subfigure}[b]{0.45\textwidth}
        \includegraphics[width=\textwidth]{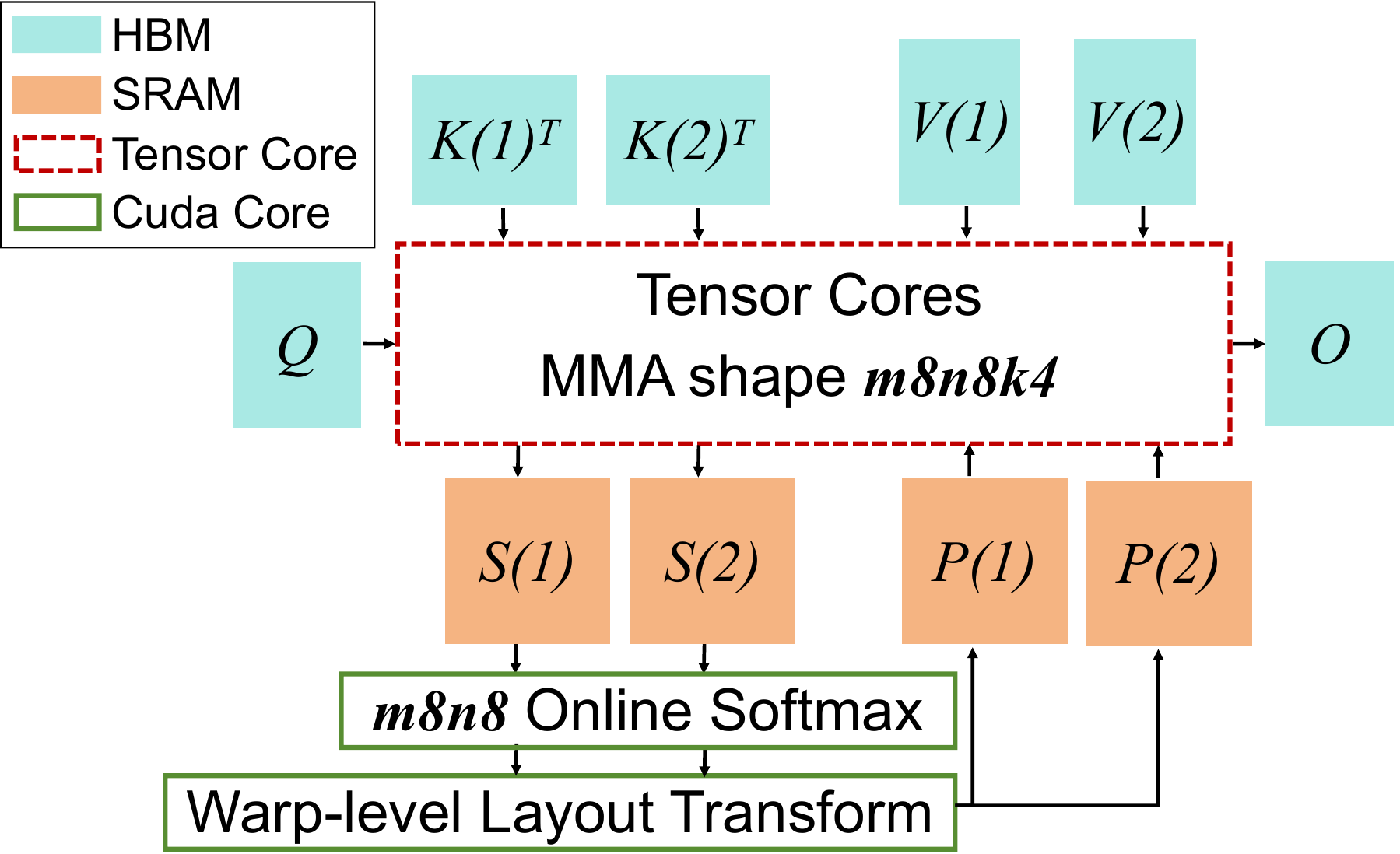}
        \caption{SparkAttention MHA-Forward implementation.}
    \end{subfigure}
    \hspace{0.0cm} 
    \begin{subfigure}[b]{0.4\textwidth}
        \includegraphics[width=\textwidth,height=4cm]{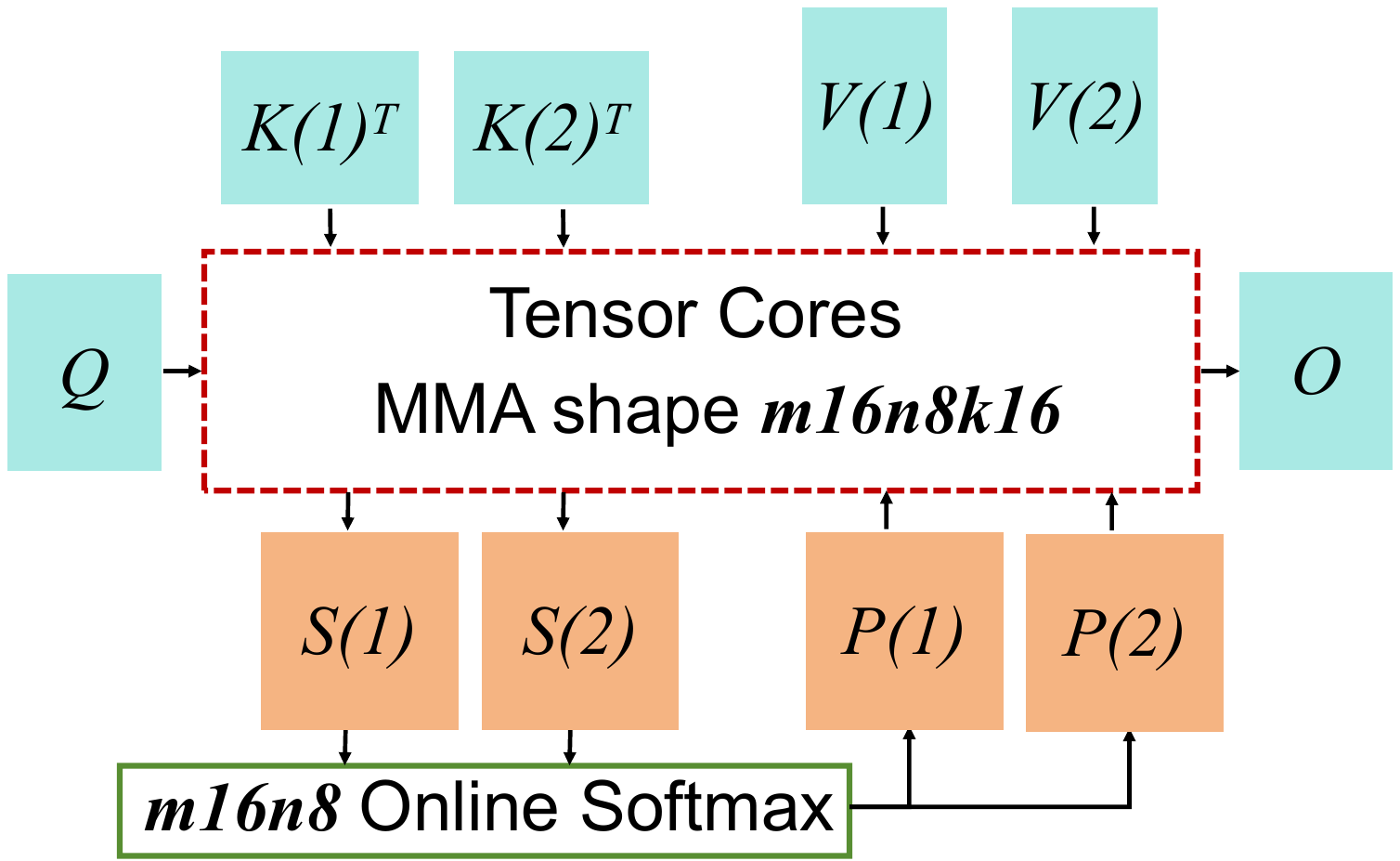}
        \vspace{2pt}
        \caption{FA2 MHA-Forward implementation.}
    \end{subfigure}
    \caption{Differences in the MHA-Forward implementations of SparkAttention and FlashAttention-2.}
    \label{fig:difference}
\end{figure*}

NVIDIA GPU consist of an array of streaming multiprocessors (SMs) and each SM contains CUDA Cores and Tensor Core Units. 
Its programming model is Single Instruction, Multiple Threads (SIMT). 
A group of threads on a GPU is called a thread block. 
Each thread block contains several warps, with each warp consisting of 32 threads. 
The scheduling unit of each SM is a warp.
TCU are only present in GPU with Volta and later architectures (e.g., Turing\cite{turing}, Ampere\cite{ampere}, Ada\cite{ada}, and Hopper\cite{hopper}). 
The theoretical peak performance of the V100 TCU (112 TFLOPS) on FP16 is $4\times$ that of CUDA Cores (28 TFLOPS).
To program on TCU, CUDA provides a warp-level API called MMA\cite{ptx}.
We can invoke it through the inline assembly.

It's important to note that V100 only supports MMA in the \textit{m8n8k4} shape. 
Assume the MMA calculation equation is $D=A \times B + C$. Here, $A$ and $B$ are called input matrices, $D$ is called the output matrix, and $C$ is called the accumulation matrix. 
In most cases, $D$ is equivalent to $C$. 

Figure \ref{fig:mma_m8n8k4} shows the differences between Volta and other architectures in MMA computation.
For the Volta architecture, each thread reads 4 elements from each of the two input matrices (MMA matrix $A$ and $B$). 
Within a warp, a single \textit{m8n8k4} calculation involves 8 threads, with each warp executing 4 such matrix computations concurrently. 
This means that 32 threads are divided into 4 MMA Computations, with each MMA Computation consisting of 8 threads. 
Each thread obtains 8 elements of the MMA matrix $C$.
For other architectures, all threads within a warp collaboratively compute the same matrix multiplication.

As shown in Figure \ref{fig:mma_computation1}, we only show the data layout requirements for MMA Computation 1, because the data layout for other computations is similar.
It should be noted that different accumulation data types will result in different data layouts for the MMA matrix $C$.

\subsection{Motivation}
To achieve higher prediction precision, most large-scale language models ~\cite{devlin2018bert,brown2020language,raffel2020exploring,liu2019roberta,yang2019xlnet,clark2020electra,radford2019language} increase their training parameters, which leads to a heavy training burden. 
GPU provides computility for large language models.
Many works accelerate the model training process through the high concurrency of GPU.

As shown in Equation \ref{equ:Transformer}, the traditional computation method of the MHA-Forward is as follows: First, read $Q$ and $K$ from HBM, compute $S$, and then write $S$ back to HBM. 
Second, read $S$ from HBM, compute $P$, and write $P$ back to HBM. 
Third, read $P$ and $V$ from HBM, compute $O$, and write $O$ back to HBM. 
Thus, we obtain the computation result of the MHA-Forward. 
In this process, we read from HBM 5 times and write to HBM 3 times. 
This computation method does not consider the I/O overhead.
Therefore, we need a more I/O-focused approach to MHA computation.

Fortunately, FA2 has become the state-of-the-art (SOTA) work for training Transformer models from an I/O perspective.
In addition, FA2 incorporates advanced optimizations for specific hardware architectures, such as NVIDIA A100\cite{ampere} and H100\cite{hopper} GPU.
But, FA2 can not run on Volta GPU due to the different shapes of MMA.

Unfortunately, the demand for computing resources is still relatively tight, thus implementing a highly efficient library to utilize the available GPU computing resources is crucial. 
This is particularly significant given the widespread availability of Volta GPU.
In the LLM scenario, the computing resources of Volta GPU have not yet been fully utilized.
But, Volta GPU is still widely deployed in clusters of major cloud service providers and remains an important provider of computing resources.
If a company already owns Volta GPUs, it is beneficial to use them for efficient model training.
So, developing acceleration libraries that enable efficient Transformer model training on Volta GPU is meaningful.
In other words, applying the FA2 algorithm on the Volta GPU is meaningful.

The specific differences between FA2 and SaprkAttention in MHA-Forward computation are shown in Figure \ref{fig:difference}. 

\section{Method}
\subsection{Overview}
\begin{figure}[htbp] \small
  \centering
  \includegraphics[width=\columnwidth]{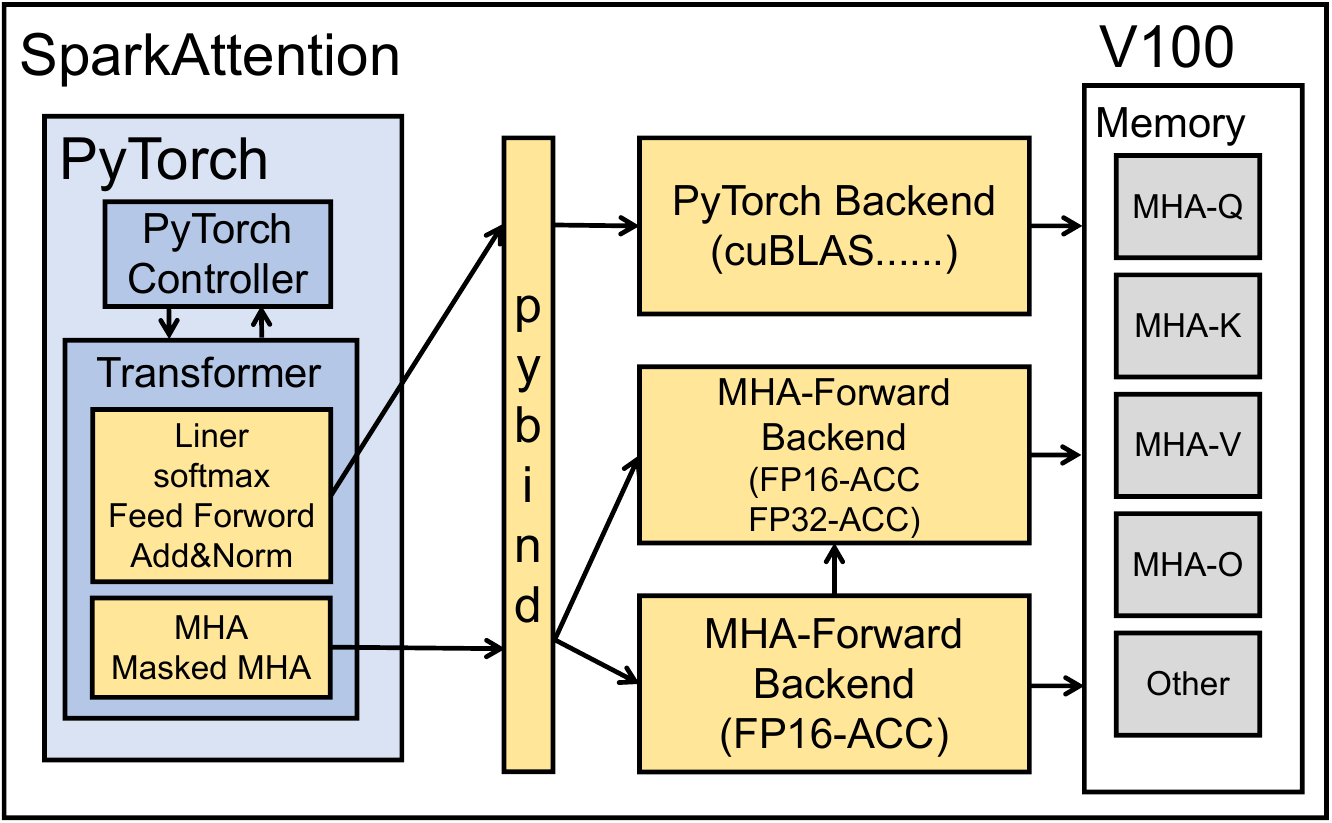}
  \caption{Overview of SparkAttention used in PyTorch.}
  \label{fig:overview}
\end{figure}

As shown in Figure \ref{fig:transformer_example}, SparkAttention completes the MHA computation using kernel fusion.
We use pybind11\cite{pybind11} to call the CUDA kernel of SparkAttention used in Python.
Figure \ref{fig:overview} shows an overview of SparkAttention used in PyTorch, which can efficiently compute the MHA on Volta GPU.
Similar to the MHA-Forward computation algorithm, we use a single CUDA kernel to complete the main calculations for the MHA-Backward.
It is important to note that during the MHA-Backward, we choose to recompute the MHA-Forward to reduce the memory usage on Volta GPU, allowing for the training of larger model.
In the CUDA kernel implementation of MHA, when the MMA matrix C uses data types FP16 or FP32 as shown in Figure \ref{fig:mma_computation1}, the corresponding kernel versions are referred to as \textbf{FP16-ACC} or \textbf{FP32-ACC}, respectively.
SparkAttention provides MHA-Forward computation in both FP16-ACC and FP32-ACC. 
However, since MHA-Backward does not require high precision, SparkAttention only offers the FP16-ACC.
Regardless of the kernel version, SparkAttention ensures that the final computation result of MHA is always in the FP16 data type.

In summary, we only focuses on optimizing the computation of MHA and does not optimize other computation of layers within the Transformer model. 
Therefore, SparkAttention is more intended to serve as an efficient MHA computation option for optimizing Volta GPU performance within other optimized training or inference libraries.

\begin{figure*}[htbp] \small
  \centering
  \includegraphics[width=0.8\textwidth]{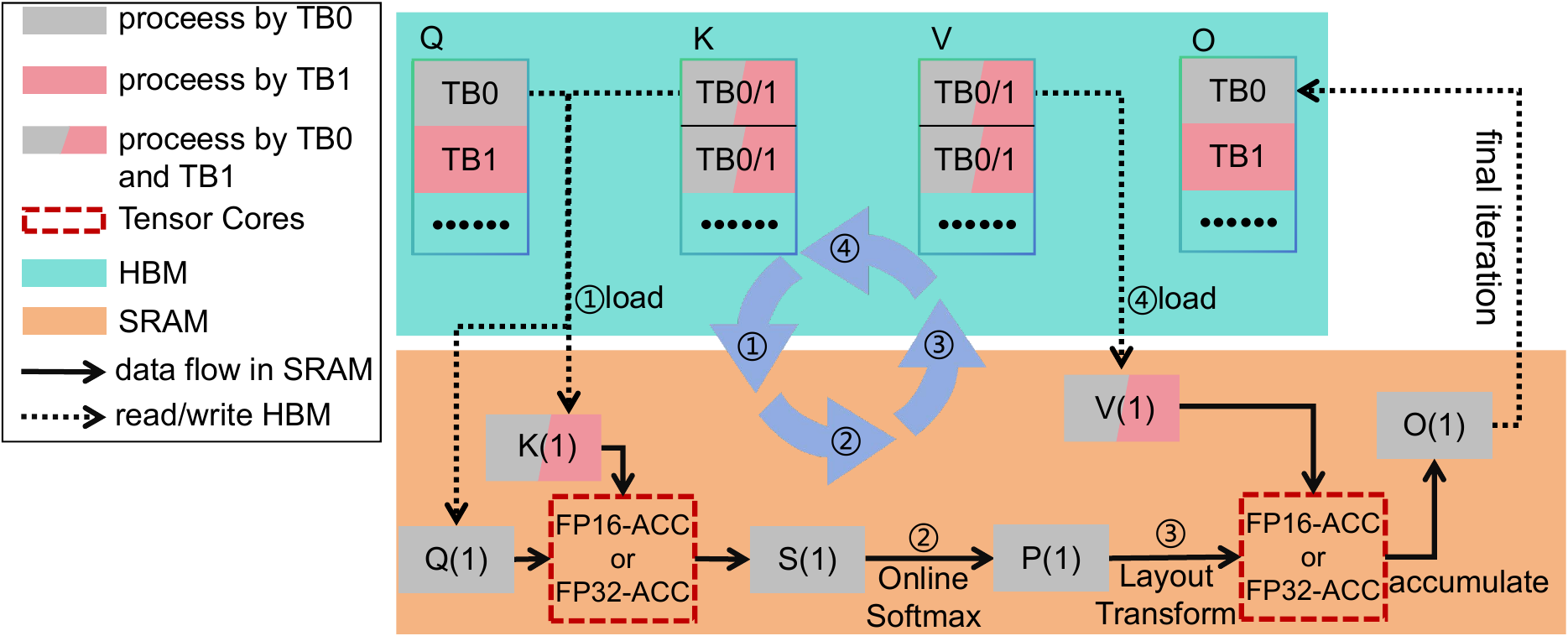}
  \caption{One iteration of thread block 0 in MHA-Forward computation of SparkAttention.}
  \label{fig:kernel_fusion}
\end{figure*}

\begin{figure}[htbp] \small
  \centering
  \includegraphics[width=\columnwidth]{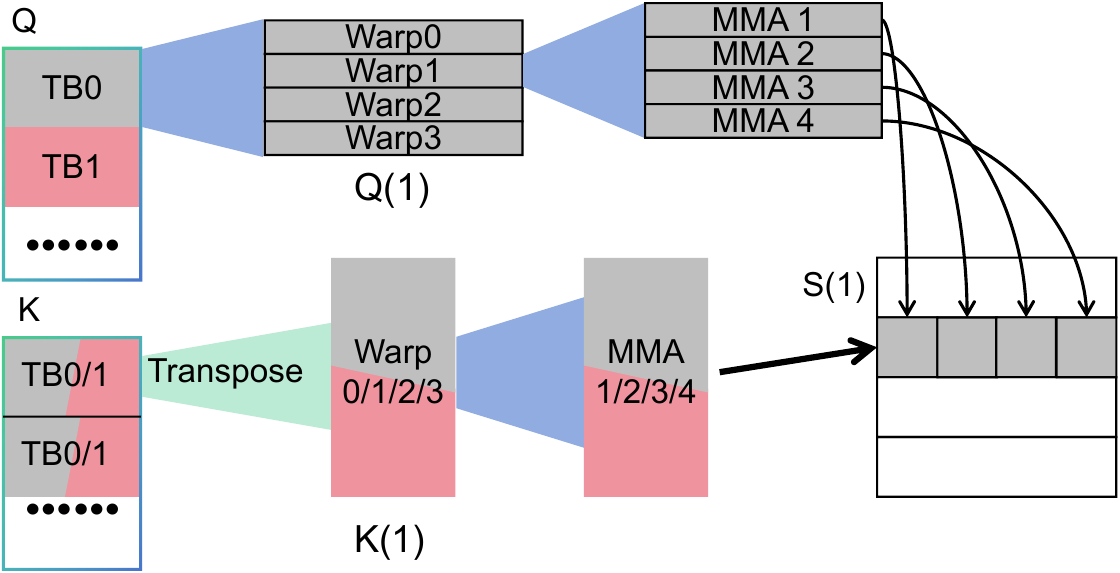}
  \caption{The computation distribution of $Q\times K$ when thread block 0 performs one iteration.}
  \label{fig:warp_arragement}
\end{figure}

\subsection{MHA-Forward Design} \label{sec:forward_design}
As shown in Figure \ref{fig:kernel_fusion}, we will use an example of one iteration of the computation process from Thread Block 0 (TB0) to illustrate the MHA-Forward computation of SparkAttention.
Blocks with the same color indicate that they are read, computed, or written back by the same TB. 
The specific calculation process for TB0 in the first iteration is as follows: 
\ding{172}Read the corresponding $Q(1)$ and $K(1)$ from HBM, and use TCUs to compute, then get the $S(1)$; 
\ding{173}Use online softmax to compute the partial softmax of $S(1)$, then get $P(1)$;
\ding{174}Transform the data layout from the MMA matrix $C$ to the data layout of the MMA matrix $A$, preparing for the subsequent multiplication with $V(1)$; 
\ding{175} Read the $V(1)$ from HBM, and use TCUs to accumulate the results into the final output $O(1)$ of TB0 in GPU SRAM. 
Following the above process, TB0 will continue iterating the calculations, and in the final iteration, the result $O(1)$ will be written back to HBM.
The computation process of other TBs are the same as TB0. 
It should be noted that each TB reads different $Q$ and computes different $O$.
By using kernel fusion to implement the MHA-Forward, we read from HBM 3 times and write to HBM 1 time. 
Compared to the traditional MHA-Forward computation described in Section 2.3, this method reduces the frequency of I/O operations and lowers the HBM overhead.

As shown in Figure \ref{fig:warp_arragement}, the arrangement of internal threads in TB0 is displayed, following the simplest layout approach. 
The calculation process of $P\times V$ is similar to that of $Q\times K$, with the only difference being the change in matrix size.
The arrangement of the MMA Computations has the potential for higher parallelism, which will be the direction of our future optimizations.

\subsubsection{Online Softmax} \label{sec:online_softmax}
Inspired by FA2, we use online softmax for computing $P$. 
We reference the online softmax calculation formula from FA2. 
First, as shown in Equation \ref{equ:softmax-normal}, we need to define some functions where $X \in \mathbb{R}^{B}$ and \textit{xi} denotes the i-th element of $X$.

\begin{figure*}[htbp] \small
  \centering
    \begin{subfigure}[b]{0.5\textwidth}
        \includegraphics[width=\textwidth]{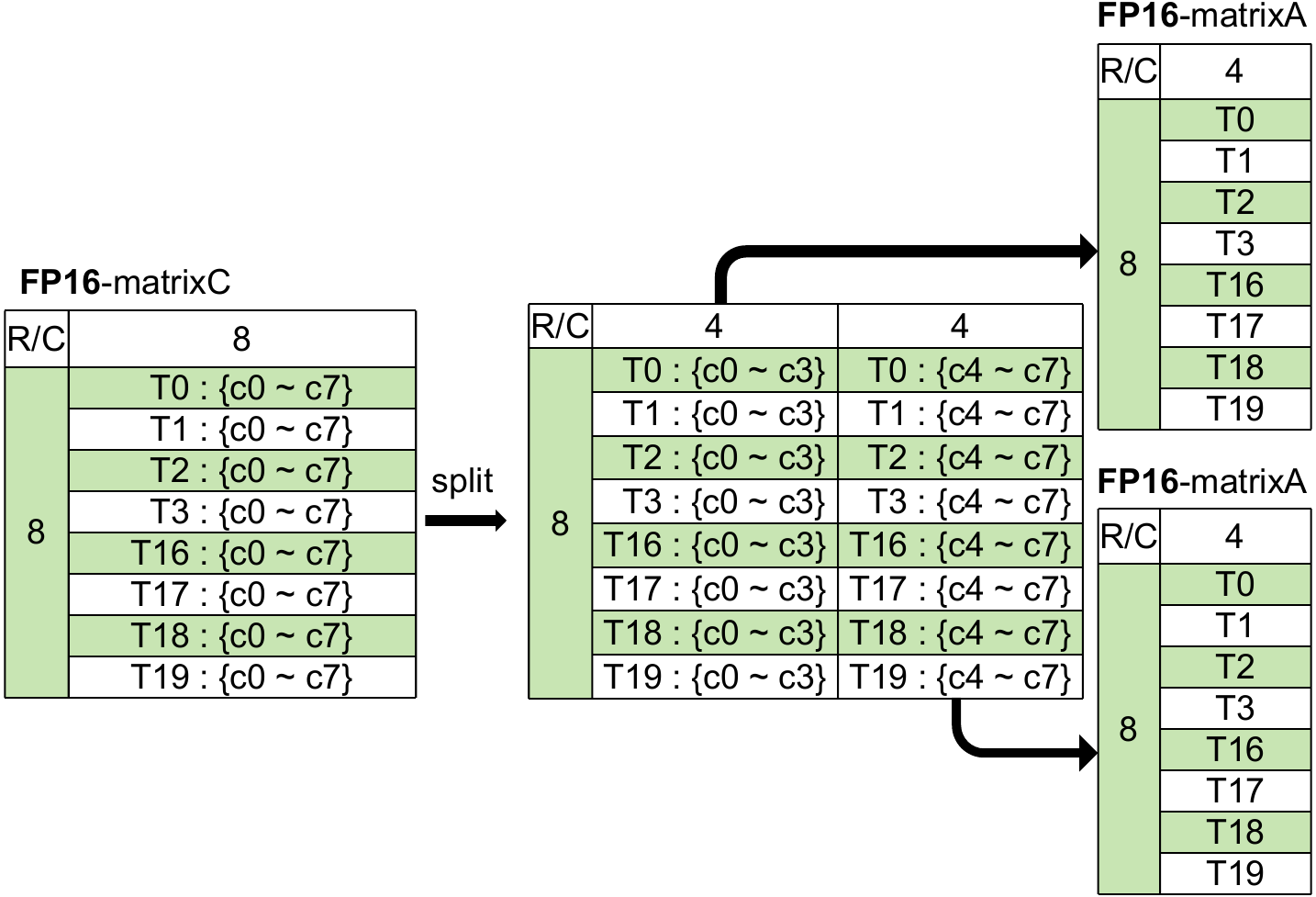}
        \vspace{3pt}
        \caption{Layout transform for FP16.}
        \label{fig:format_conversion-a}
    \end{subfigure}
    \hspace{0.0cm} 
    \begin{subfigure}[b]{0.48\textwidth}
        \includegraphics[width=\textwidth]{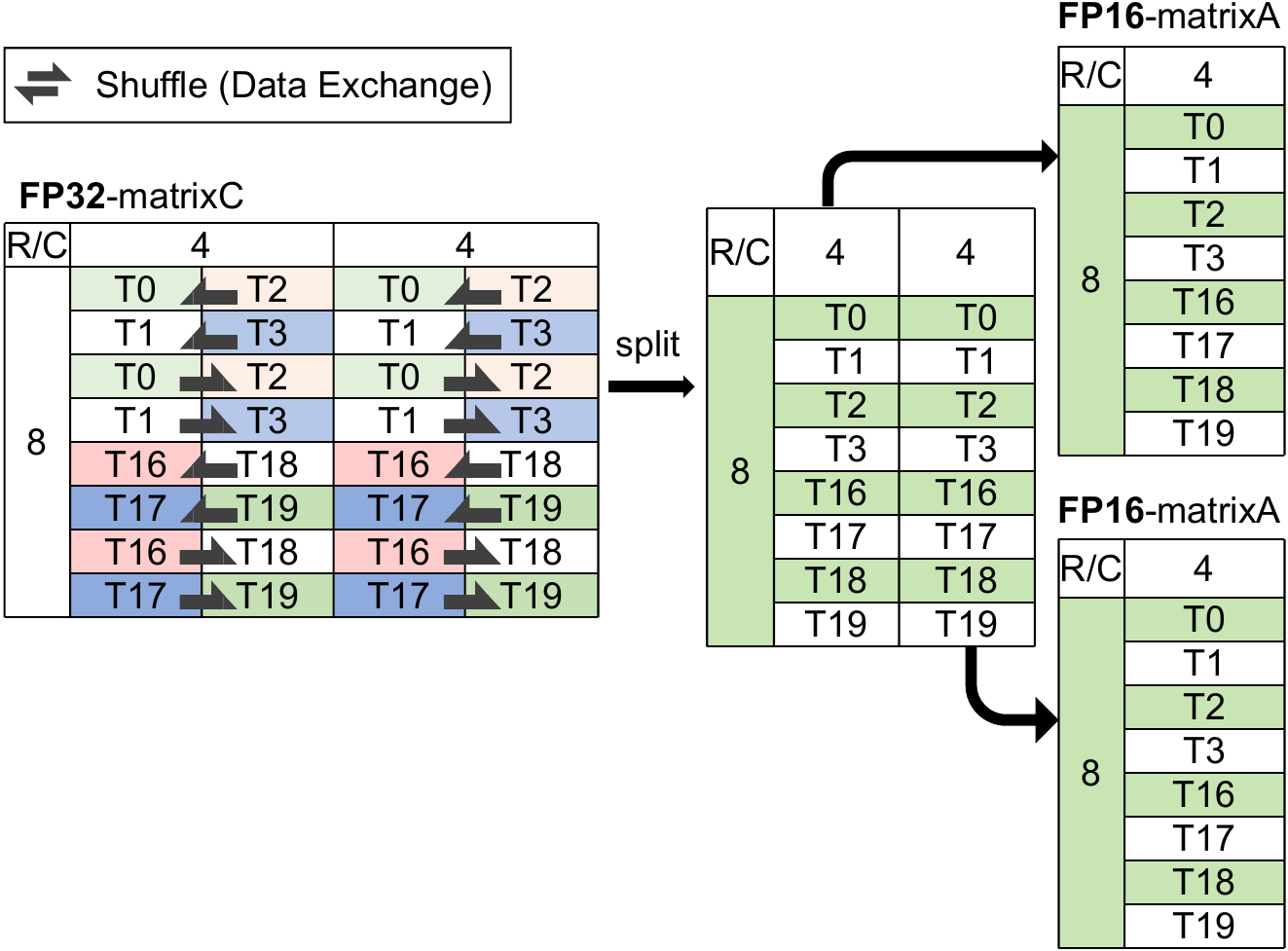}
        \caption{Layout transform for FP32.}
        \label{fig:format_conversion-b}
    \end{subfigure}
  \caption{Layout transform of matrix C to match the layout of matrix A in MMA (m8n8k4).}
  \label{fig:format_conversion}
\end{figure*}

\begin{equation} \small \label{equ:softmax-normal}
\begin{gathered}
    m(X):=\max_{i}(xi)\\
    f(X):=[e^{(x_1 - m(X))}~...~e^{(x_B - m(X))}]\\
    l(X):=\sum_{i=1} x_i\\
    softmax(X) := \frac{f(X)}{l(X)}\\
\end{gathered}
\end{equation}

According to the above function, we can decompose the large softmax with scaling. 
For example, $X^1,X^2 \in \mathbb{R}^{B}, X=[X^1X^2] \in \mathbb{R}^{2B}$. 
If we want to compute the softmax of \textit{X}, we can use the functions in Equation \ref{equ:softmax-itreation}.

\begin{equation} \small \label{equ:softmax-itreation}
\begin{gathered}
    m(X)=m([X^1~~X^2])=max(m(X^1), m(X^2))\\
    f(X)=[e^{m(X^1) - m(X)}f(X^1)~~e^{m(X^2) - m(X)}f(X^2)]\\
    l(X)=e^{m(X^1) - m(X)}l(X^1) + e^{m(X^2) - m(X)}l(X^2)\\
    softmax(X)=\frac{f(X)}{l(X)}\\
\end{gathered}
\end{equation}

\begin{figure*}[htbp] \small
  \centering
  \includegraphics[width=0.7\textwidth]{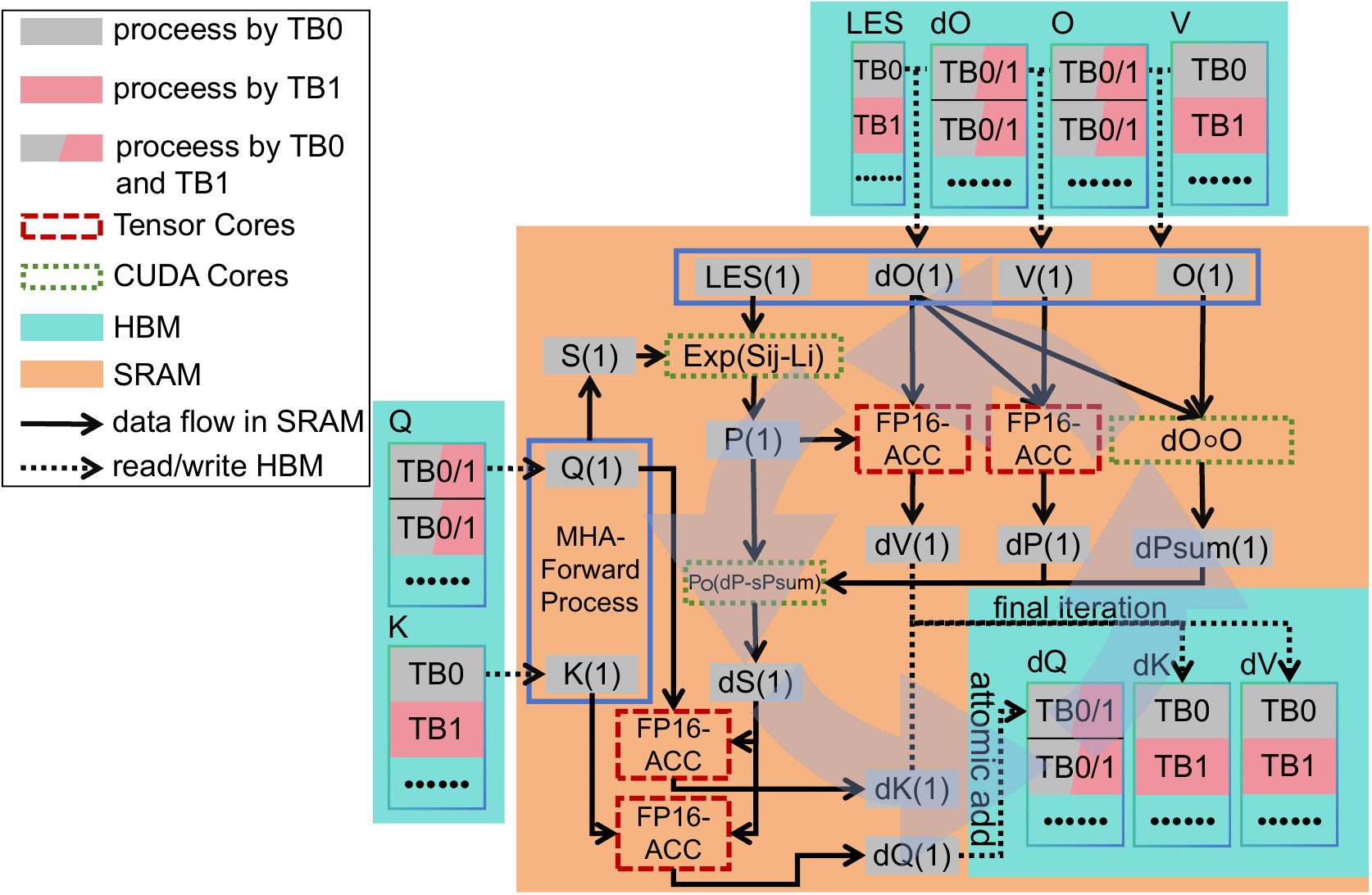}
  \caption{One iteration of thread block 0 in MHA-Backward computation of SparkAttention.}
  \label{fig:kernel_fusion_backward}
\end{figure*}

From the above equation, it can be seen that to iteratively compute the softmax, it is necessary to keep track of the maximum value and perform summation. 
It's important to note that when using online softmax and kernel fusion, during each iteration, each thread needs to restore the previously accumulated matrix $O$ by multiplying it with the current row's $e^{m(pre) - m(cur)}$, and then accumulate again to ensure accuracy in the computation. 
$m(pre)$ refers to the maximum value of each row from the previous iteration, and $m(cur)$ refers to the maximum value of each row in the current iteration.

\textbf{FP16-ACC Online Softmax.}
To obtain the maximum value and the sum, we only need to perform accumulation and find the maximum value within each thread.
When performing softmax calculations at FP16, we need to convert it to FP32 to ensure that the softmax computation does not result in errors or overflow due to precision limitations.
Specifically, we need to convert the current output matrix $S$ and the final result matrix $O$ (need to multiply $e^{m(pre) - m(cur)}$) from FP16 to FP32 to perform the online softmax calculation.
However, our subsequent experimental tests demonstrate that these data type conversions incur a certain level of overhead, which can sometimes be even higher than that of the shuffle operation used in FP32-ACC online softmax.
We attempted calculations without performing data type conversion, which resulted in an average absolute error of one-tenth.
Given that most computed values are below 1, we consider this one-tenth absolute error to be a significant precision issue. 
Therefore, data type conversion is necessary. 

\textbf{FP32-ACC Online Softmax.}
The advantage of using FP32-ACC is that, during softmax computation, there is no need for data type conversion, naturally ensuring the accuracy of softmax calculations.
In other words, FP32-ACC online softmax requires two fewer data type conversions than FP16-ACC.
However, we need to use shuffle operations to exchange thread data in order to obtain the maximum value and sum of the current row.
As shown in Figure \ref{fig:format_conversion-b}, during the data exchange process, we only need to change each one-way arrow to a two-way arrow.
Specifically, in the first row, Thread 0 not only needs to receive data from Thread 2 but also needs to send its own data to Thread 2.

\subsubsection{Warp-Level Layout Transform}\label{sec:Warp-Level_Layout_Transform}
To ensure the correctness of two-stage matrix multiplication, we need to transform the data storage layout of MMA matrix C into the data storage layout of MMA matrix A. 
This ensures the accuracy of the subsequent multiplication with the matrix V as shown in Figure \ref{fig:kernel_fusion}\ding{174}. 
The different data types of MMA matrix C will lead to different layout transformations. 

\textbf{FP16-ACC Layout Transform.}
As shown in Figure \ref{fig:format_conversion-a}, for FP16-ACC, we only need to split the computed result, and the segmented results directly satisfy the data layout requirements of MMA matrix A. 
In other ward, the only thing we need to do is to operate on registers of the same thread to obtain the data layout required for the next matrix computation. 

\textbf{FP32-ACC Layout Transform.}
As shown in Figure \ref{fig:format_conversion-b}, we need to perform data exchange across different threads. 
Specifically, we use $shfl\_xor\_sync(2)$ for data exchange between threads whose thread IDs XOR to 2. 
For example, the 0th row requires transferring data from thread 2 to thread 0, and the 2nd row requires transferring data from thread 0 to thread 2. 
However, we need to have a data type conversion from FP32 to FP16 in the next MMA calculation to meet the data type requirements of MMA matrix A.
In addition, a similar layout transform (just the threads are different) occurs for other MMA Computations as shown in Figure \ref{fig:format_conversion-b}.

\subsection{MHA-Backward Design}
During the MHA-Backward computation, we choose to recompute the MHA-Forward to save memory overhead.
Similar to the MHA-Forward computation, we use kernel fusion to implement the calculations during the MHA-Backward pass.

As shown in Figure \ref{fig:kernel_fusion_backward}, we use one iteration of the computation process from Thread Block 0 (TB0) to illustrate the MHA-Backward computation of SparkAttention. 
In the Figure \ref{fig:kernel_fusion_backward}, $LES$ represents the maximum value of the current row in softmax recorded during the MHA-Forward computation. 
$dPsum$ is the component needed for calculating $dS$ at the mathematical level.
$Exp(S_{ij}-L_i)$ represents the element in the i-th row and j-th column of matrix $S$ minus the maximum value recorded in the i-th row of $LES$.
$\circ$ denotes element-wise multiplication at corresponding positions.
It should be noted that we are only focusing on the computational aspects. 
For the mathematical derivation of MHA-Backward computation, we will not provide a proof here.
The specific MHA-Backward computation is shown in Equation \ref{equ:backward}.

\begin{equation} \small \label{equ:backward}
\begin{gathered}
     dV=P^TdO \in \mathbb{R}^{N \times d}\\
     dP=dOV^T \in \mathbb{R}^{N \times N}\\
     dS=dsoftmax(dP) \in \mathbb{R}^{N \times N}\\
     dQ=dSK \in \mathbb{R}^{N \times d}\\
     dK=QdS^T \in \mathbb{R}^{N \times d}\\
\end{gathered}
\end{equation}

Drawing from the MHA-Forward computation method, we also implement the entire MHA-Backward computation using a single CUDA kernel.
It is important to note that during each iterative computation, we only accumulate calculations for $dK$ and $dV$, while $dQ$ needs to be accumulated into HBW using atomic add each time.
In other words, different TBs will only compute a portion of the $dK$ and $dV$ matrices, but
some TBs will compute the same portion of the $dQ$ matrix.
\begin{figure*}[t] \small
    \centering
    \begin{subfigure}[b]{0.24\textwidth}
        \includegraphics[width=\textwidth]{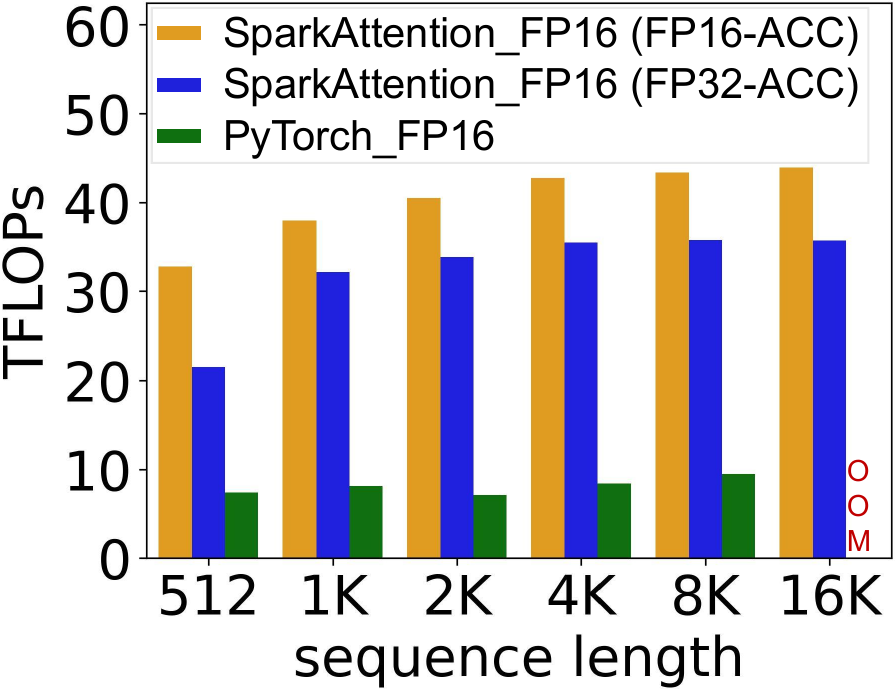}
        \caption{head-dimension=64 \\ causal mask=False}
        \label{fig:subfig_forward-a}
    \end{subfigure}
    \hspace{0.0cm} 
    \begin{subfigure}[b]{0.24\textwidth}
        \includegraphics[width=\textwidth]{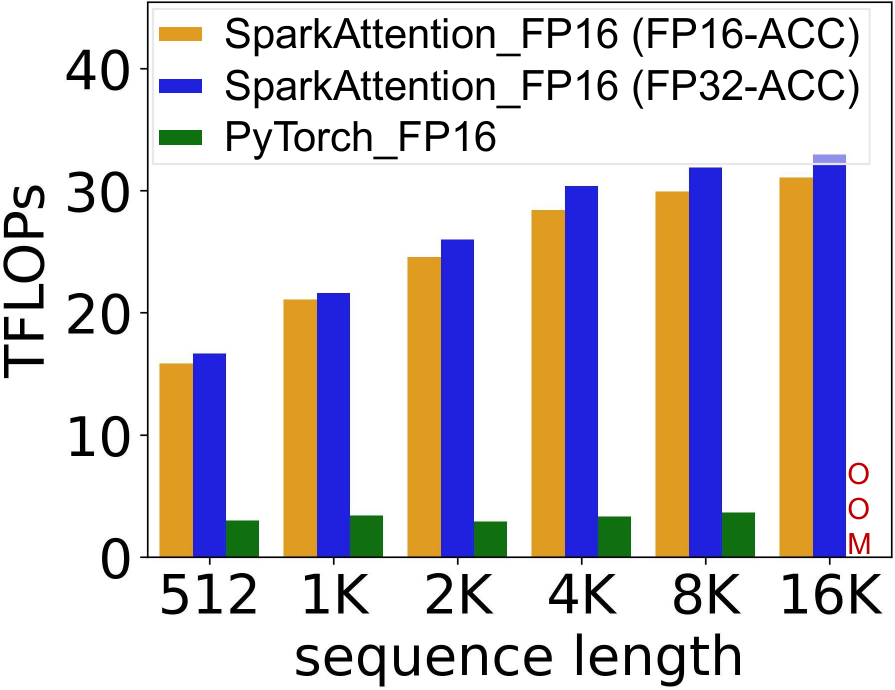}
        \caption{head-dimension=64 \\ causal mask=True}
        \label{fig:subfig_forward-b}
    \end{subfigure}
    \hspace{0.0cm}
    \begin{subfigure}[b]{0.24\textwidth}
        \includegraphics[width=\textwidth]{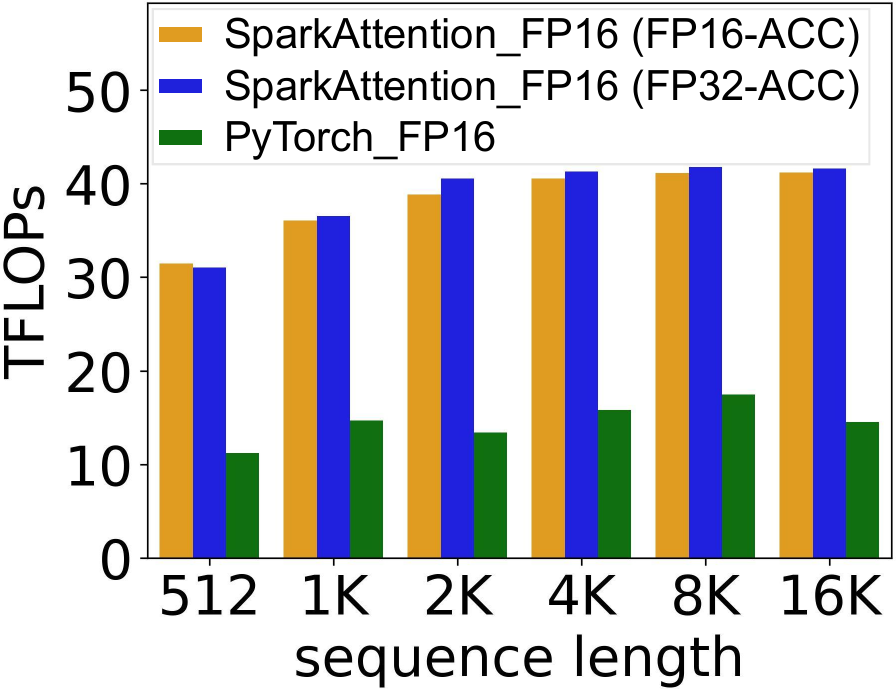}
        \caption{head-dimension=128 \\ causal mask=False}
        \label{fig:subfig_forward-c}
    \end{subfigure}
    \hspace{0.0cm}
    \begin{subfigure}[b]{0.24\textwidth}
        \includegraphics[width=\textwidth]{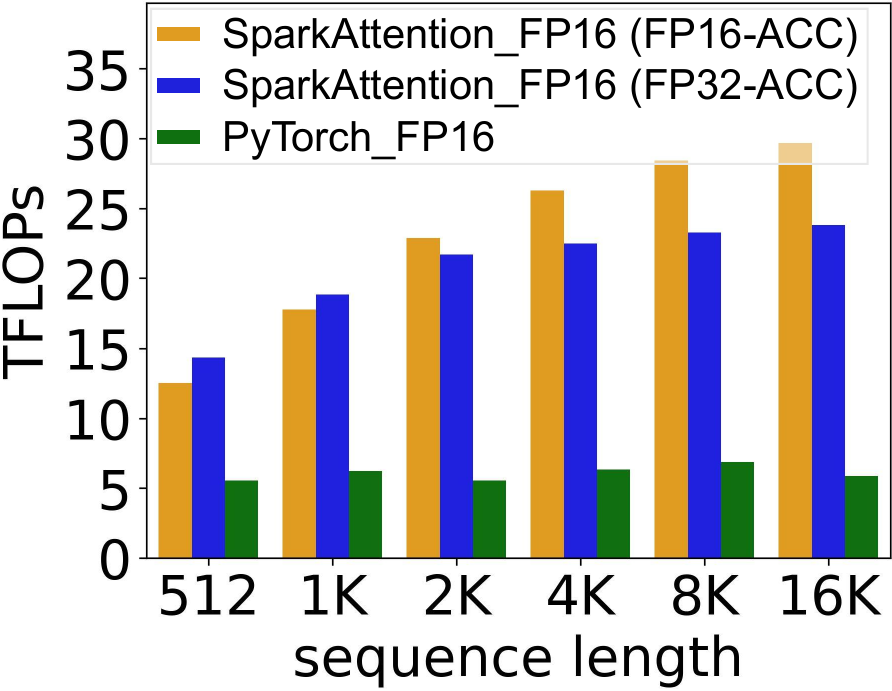}
        \caption{head-dimension=128 \\ causal mask=True}
        \label{fig:subfig_forward-d}
    \end{subfigure}
    \caption{The performance of SparkAttention MHA-Forward computation on V100.}
    \label{fig:subfig_forward}
\end{figure*}

\section{Evaluation}\label{sec5}

\subsection{Experimental Setup} \label{sec:setup}

\textbf{System Configuration.}
Our experimental environment is as follows: the operation system version is Ubuntu 20.04, the CUDA version is 12.1, the PyTorch version is 2.5.1, and the Python version is 3.9.16.
In addition, we evaluate SparkAttention on an NVIDIA V100 GPU (V100-SXM2-32GB GPU). 
The V100 GPU has 80 SMs, each with a total of 128KB configurable L1 cache and 64KB registers.

\noindent\textbf{Baseline.}
Our work focuses on Transformer model training. 
We compare the performance of MHA computations with PyTorch\_FP16. 
In the end-to-end tests, we only compare the Encoder-Forward computation with FasterTransformer (version 5.3), ByteTransformer (version 1.0), and TurboTransformer (version 0.5.1), as these works are focused on accelerating the inference phase of Transformer. 
The reason we do not compare with FA2 is that it does not support running on Volta GPU, and due to hardware limitations, SparkAttention cannot run efficiently on GPU outside the Volta architecture.

\noindent\textbf{Hyperparameter.}
We evaluated the two most common head dimensions: 64 and 128. 
To maintain a constant hidden dimension of 2048, we set the number of heads to 2048$\div$head-dimension. 
To demonstrate the advantage of SparkAttention in long sequence scenarios, we tested sequence lengths of 512, 1024, 2048, 4096, and 16384, and set the batch size to 16384$\div$sequence-length.
To compare the differences brought by using causal masking, we set causal masking to either True or False.
Since dropout is needed to improve the model's generalization ability in most practical applications, we set the dropout rate to 0.1 in our tests. 
We use the above settings for configuring SparkAttention and all baselines.

\noindent\textbf{Dataset.}
Since we only test performance and computational accuracy, we use random numbers as the dataset.
If not explicitly stated (like \_FP32), all experiments in this section use FP16 data type as input, because SparkAttention only supports FP16.

\begin{figure*}[t] \small
    \centering
    \begin{subfigure}[b]{0.24\textwidth}
        \includegraphics[width=\textwidth]{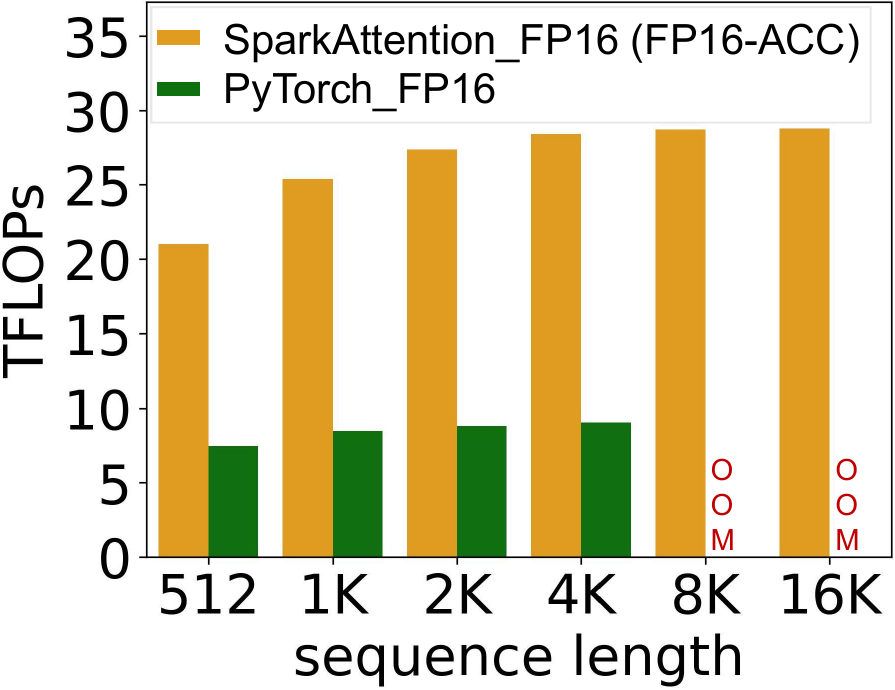}
        \caption{head-dimension=64 \\ causal mask=False}
        \label{fig:subfig_backward-a}
    \end{subfigure}
    \hspace{0.0cm} 
    \begin{subfigure}[b]{0.24\textwidth}
        \includegraphics[width=\textwidth]{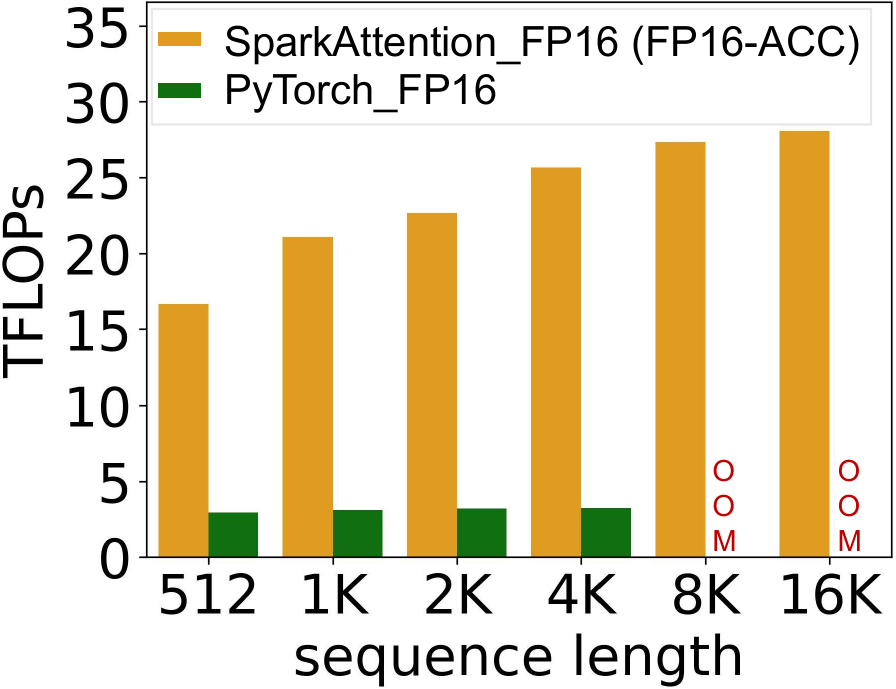}
        \caption{head-dimension=64 \\ causal mask=True}
        \label{fig:subfig_backward-b}
    \end{subfigure}
    \hspace{0.0cm}
    \begin{subfigure}[b]{0.24\textwidth}
        \includegraphics[width=\textwidth]{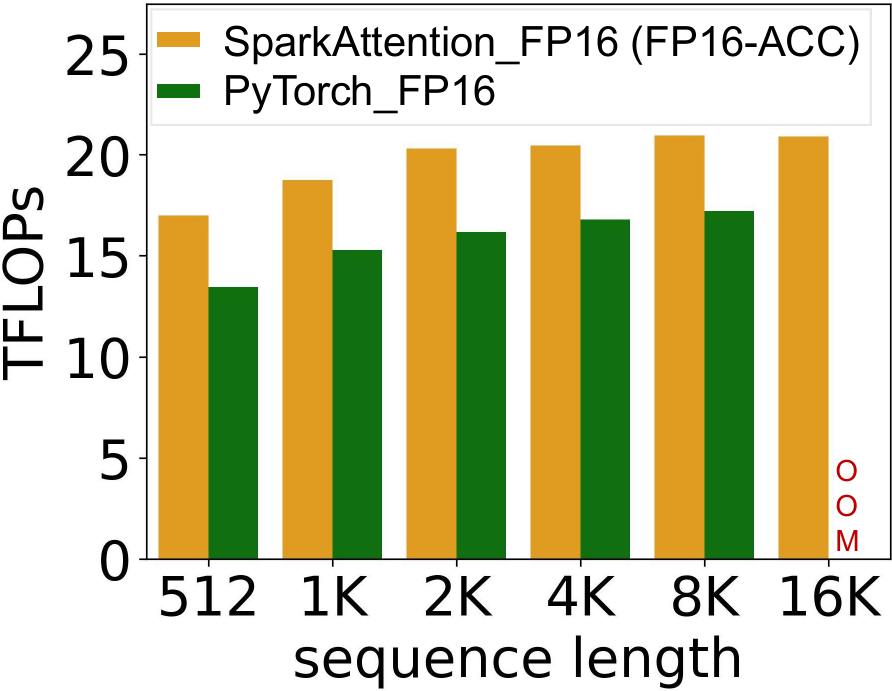}
        \caption{head-dimension=128 \\ causal mask=False}
        \label{fig:subfig_backward-c}
    \end{subfigure}
    \hspace{0.0cm}
    \begin{subfigure}[b]{0.24\textwidth}
        \includegraphics[width=\textwidth]{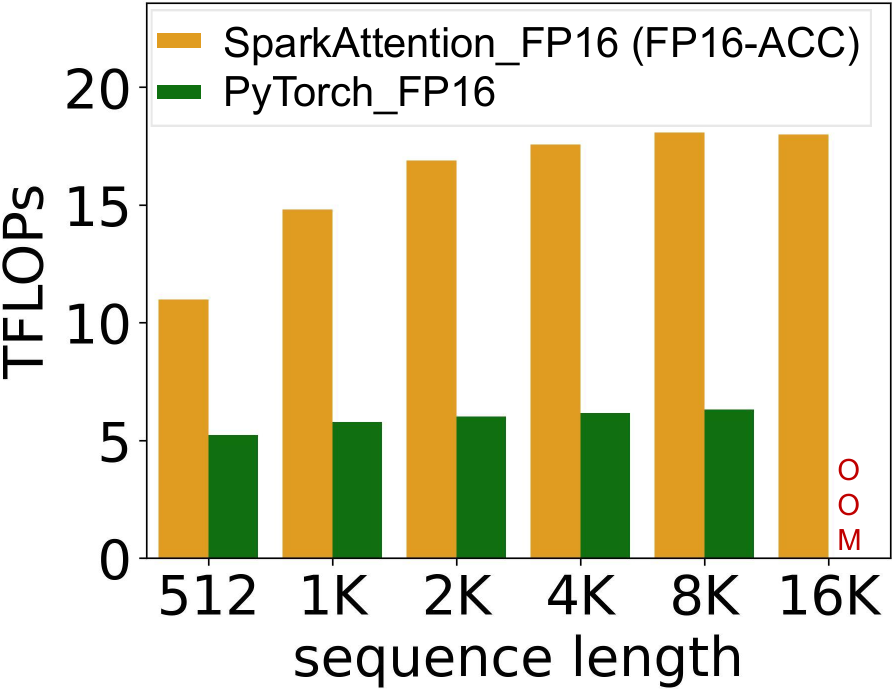}
        \caption{head-dimension=128 \\ causal mask=True}
        \label{fig:subfig_backward-d}
    \end{subfigure}
    \caption{The performance of SparkAttention MHA-Backward computation on V100.}
    \label{fig:subfig_backward}
\end{figure*}

\begin{figure*}[t] \small
    \centering
    \begin{subfigure}[b]{0.45\textwidth}
        \includegraphics[width=\textwidth]{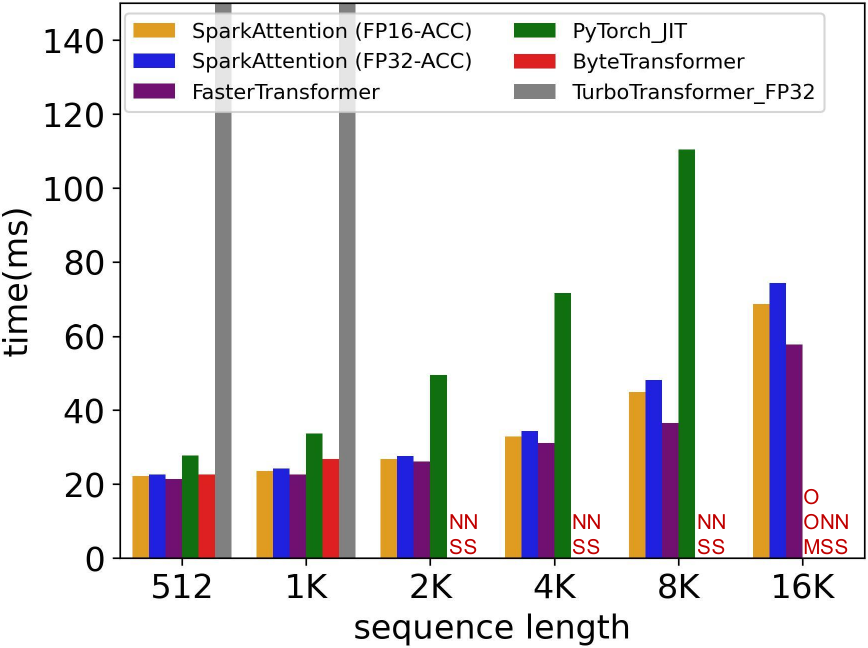}
        \caption{head-dimension=64 causal mask=False}
        \label{fig:subfig_end2end-a}
    \end{subfigure}
    \hspace{1cm} 
    \begin{subfigure}[b]{0.45\textwidth}
        \includegraphics[width=\textwidth]{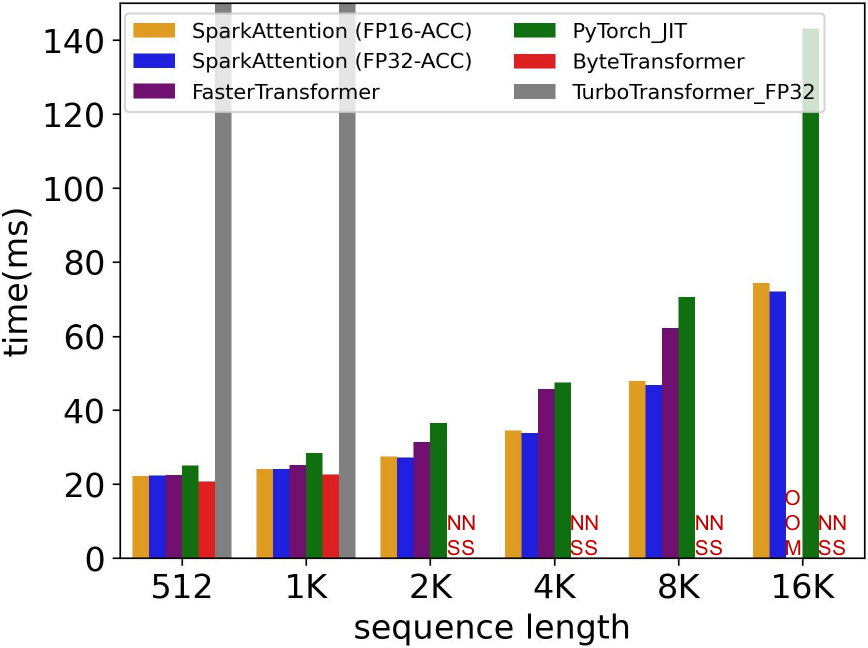}
        \caption{head-dimension=128 causal mask=False}
        \label{fig:subfig_end2end-b}
    \end{subfigure}
    \caption{The performance of SparkAttention in traditional Encoder-Forward experimental results on V100 (OOM: out of memory; NS: not support).}
    \label{fig:subfig_end2end}
\end{figure*}

\subsection{Effectiveness of SparkAttention}

\subsubsection{Effectiveness of MHA-Forward} \label{sec:MHA-Forward_test}
For MHA-Forward computation, we provide FP16-ACC and FP32-ACC.
The reason we do this is that using FP16-ACC sometimes can achieve better performance than FP32-ACC due to a trade-off between data type conversion and shuffle operation.
However, FP16-ACC may results in a loss of computational precision.
It is also worth noting that FP16-ACC and FP32-ACC are identical in terms of pure performance. 
This means that the performance of FP16-ACC is not necessarily better than that of FP32-ACC. 
The performance differences are actually influenced by data type conversion in FP16-ACC and shuffle operations in FP32-ACC.

The experimental results are shown in Figure \ref{fig:subfig_forward}.
Specifically, compared to PyTorch\_FP16, the FP16-ACC achieves an average speedup of 4.66$\times$ (up to 8.56$\times$) while the FP32-ACC achieves an average speedup of 4.44$\times$ (up to 9.17$\times$).
We can also observe that as the sequence length increases, PyTorch\_FP16 gradually reveals out-of-memory issues.
However, even with a sequence length of 16,384, SparkAttention still achieves excellent performance.
At the same time, as the sequence length increases, SparkAttention can better utilize the performance of V100 GPU.
Also, when the head dimension is large, the computation per head increases, allowing the computational capabilities of GPUs to be fully utilized. 
This explains why PyTorch\_FP16 achieves higher TFLOPs when the head dimension is 128.
It is important to note that when “casual mask=True” is used to compute TFLOPs, the computational workload is reduced by half under the same configuration.

\subsubsection{Effectiveness of MHA-Backward}
In the MHA-Backward, we only implemented the FP16-ACC. 
We applly the same dropout logic as in the MHA-Forward process to obtain consistent dropout results.
Figure \ref{fig:subfig_backward} shows the performance of our experimental results. 
We can see SparkAttention MHA-Backward implementation significantly outperforms implementation of PyTorch\_FP16.
At the same time, SparkAttention MHA-Backward computation is able to run on limited GPU memory even when handling long sequences.
Specifically, compared to PyTorch\_FP16, SparkAttention achieves an average speedup of 3.44$\times$ (up to 7.91$\times$).

\subsubsection{MHA Accuracy} \label{subsec:mha_accuracy}
\textbf{MHA-Forward.}
Using the PyTorch\_FP32 as the benchmark, We test the accuracy of the FP32-ACC and FP16-ACC. 
In FP32-ACC, the results show that the average relative error is 0.035\%, and the average absolute error is 0.0019\%.
In FP16-ACC, the results show that the average relative error is 0.76\%, and the average absolute error is 0.01\%.

\noindent\textbf{MHA-Backward.}
We also test the accuracy of the FP16-ACC using PyTorch\_FP32.
The results show that the average relative error is 0.23\%, and the average absolute error is 0.0022\%.

To demonstrate that our computational error is acceptable, we also calculate the computational error of PyTorch\_FP16 compared to PyTorch\_FP32.
Specifically, for MHA-Forward computation, the average relative error of PyTorch\_FP16 is 0.065\%, and the average absolute error is 0.0048\%. For MHA-Backward computation, the average relative error is 0.40\%, and the average absolute error is 0.0053\%.

In summary, the computation accuracy of SparkAttention is within an acceptable range.

\subsubsection{Effectiveness of End-to-End} \label{subsec:EndtoEnd}
In the End-to-End experiments, we select the traditional Encoder model architecture for testing.
All baselines include only a single Encoder layer, and each experiment is repeated 30 times, with the average value taken as the final result.

The specific experimental results, as shown in Figure \ref{fig:subfig_end2end}, demonstrate that SparkAttention achieves significantly lower execution time compared to PyTorch\_JIT.
To make a fair comparison, we only replace the MHA-Forward computation in PyTorch\_JIT with the MHA-Forward computation from SparkAttention, and use this as the End-to-End implementation for SparkAttention. 
This method allows us to observe the benefits brought by using SparkAttention MHA-Forward through the control variable method.
Specifically, compared to PyTorch\_JIT, SparkAttention achieves an average speedup of 1.80$\times$ (up to 2.46$\times$).

It is important to note that SparkAttention is designed to optimize MHA computation. 
However, due to the customized implementations of TurboTransformer, FasterTransformer, and ByteTransformer, replacing their MHA-Forward computations with SparkAttention MHA-Forward is extremely challenging. 
As a result, our comparison with other baselines inherently has disadvantages.
Specifically, ByteTransformer and TurboTransformer are unable to run on long sequences when using different head dimensions.
This further highlights the advantages of SparkAttention in handling long sequence computations (lower HBM overhead).
When the head dimension is 128, SparkAttention outperforms FasterTransformer. 
However, with the head dimension of 64, the situation is reversed. 
This is because, excluding the computation of MHA-Forward, FasterTransformer leverages techniques such as layer fusion and automatic tuning of MatMul kernels when performing other computations.

In summary, from the I/O perspective, SparkAttention provides an efficient MHA computation process for long sequences on Volta GPU. 
When other Transformer acceleration libraries aim to optimize for Volta GPU, we hope SparkAttention can serve as a preferred choice for MHA computation.

\section{Related Work} \label{sec6}

\textbf{Model architecture.} Works such as Reformer\cite{kitaev2020reformer}, Longformer\cite{beltagy2020longformer}, xformers\cite{xFormers2022}, and Big Bird\cite{zaheer2020big} utilize sparse attention to transform the training computation complexity from the previous $O(N^2)$ to $O(N)$. 
Without compromising model accuracy, this approach alleviates the issue of excessively long training times to some extent.  
These methods transform two matrix multiplications into one Sampled-Dense-Dense Matrix Multiplication (SDDMM) and one Sparse Matrix-Matrix Multiplication (SpMM) with high sparsity, aiming to achieve shorter computation times. 
These works have experimentally demonstrated that approximate attention does not significantly impact the accuracy of the models.
Sparse attention methods only consider computational efficiency. 
However, in most training processes, I/O is the bottleneck for overall runtime speed. 
Besides, in some implementations (Big Bird in Hugging Face\cite{wolf-etal-2020-transformers}) of Sparse attention models, Sparse attention computation is achieved by calling multiple GeMM operations instead of using a single SDDMM operator. 
This inefficient implementation often results in less noticeable improvements in training time.

In addition, Ying et al.\cite{ying2021lazyformer} proposed Lazyformer that improves computational efficiency by reducing the frequency of computing self-attention distributions. 
In each Lazy block, only the first layer computes the self-attention distribution, which is then reused by subsequent layers, significantly reducing computational costs. 
Fan et al.\cite{fan2019reducing} proposed using structured dropout techniques to dynamically reduce the depth of Transformer models. 
Zhang et al.\cite{zhang2020accelerating} proposed using progressive layer dropping techniques to gradually reduce the number of layers in the model during training, thereby reducing computational costs and accelerating training speed without compromising model performance and accuracy. 
Choromanski et al.\cite{choromanski2020rethinking} use low-rank approximation methods to replace traditional self-attention mechanisms.

\noindent\textbf{Training strategy.} Liu et al.\cite{liu2019variance} proposed RAdam and demonstrated that the warm-up strategy for learning rates can effectively reduce variance, thereby stabilizing the training process, accelerating convergence, and improving generalization performance. 
Gong et al.\cite{gong2019efficient} and Li et al.\cite{li2020shallow} proposed accelerating training by transferring knowledge between shallow and deep models, and progressively applying stacking.

\noindent\textbf{Optimization algorithms.} Yao et al.\cite{yao2021adahessian} proposed the adaptive second-order optimization algorithm AdaHessian, which improves the optimization process by dynamically estimating the Hessian matrix of the loss function.

\noindent\textbf{Data precision.} Zhang et al.\cite{zhang2020fixed} proposed a novel training approach that addresses significant accuracy loss issues caused by direct quantization in deep neural networks, by applying layer-wise precision-adaptive quantization. 
Sun et al.\cite{sun2019hybrid} use a hybrid 8-bit floating point for training and inference of deep neural networks.

In addition to the above, Nvidia Apex\cite{nvidia2020apex} supports automatic mixed precision training and distributed training without compromising computational accuracy;
DeepSpeed\cite{rasley2020deepspeed} integrates small kernels such as \textit{LayerNorm} and \textit{Softmax} into the encoder; 
TurboTransformers\cite{fang2021turbotransformers}, Orca\cite{yu2022orca}, and pagedattention\cite{kwon2023efficient} are mainly used to serve the inference phase of Transformers;
TVM\cite{chen2018tvm} automatically generates kernels through compilation and achieves graph-level and operator-level optimization in deep learning computations.

\section{Conclusion}\label{sec7}

In this paper, we propose SparkAttention, an acceleration library focuses on efficiently computing MHA on the Volta GPU architecture. 
SparkAttention addresses the current issue where Volta GPU, despite being widely used, lacks efficient algorithms for fast MHA computation, thereby underutilizing the computational power that Volta GPU can offer during Transformer model training.
SparkAttention leverages TCU and kernel fusion to reduce HBM accesses and overhead.
Our experiments conduct on the V100 show that SparkAttention significantly outperforms the PyTorch\_FP16 implementation in both MHA-Forward and MHA-Backward. 
In the MHA-Forward process, SparkAttention achieves an average speedup of 4.55$\times$ (up to 9.17$\times$). 
In the MHA-backward process, SparkAttention achieves an average speedup of 3.44$\times$ (up to 7.91$\times$). 
According to the End-to-End experimental results, SparkAttention achieves an average speedup of 1.80$\times$ compared to the PyTorch\_JIT implementation. 
Although this work is implemented on NVIDIA Volta GPU, we foresee that our basic approach can be easily adapted to AMD GPU and Huawei Ascend NPU. 
We also plan to further optimize SparkAttention to support more heterogeneous accelerators that current MHA algorithms do not support.

\section{Acknowledgement}
This project was supported by the National Science and Technology Major Project (2023ZD0120502), the National Natural Science Foundation of China under Grant No. 62372055, the Fundamental Research Funds for the Central Universities, the fund of Laboratory for Advanced Computing and Intelligence Engineering.

\section{Declaration}
\textbf{Conflict of interest. }On behalf of all authors, the corresponding author states that there is no conflict of interest.

\bibliographystyle{IEEEtran}
\bibliography{IEEEabrv, references}

\begin{thebibliography}{10}
\providecommand{\url}[1]{#1}
\csname url@samestyle\endcsname
\providecommand{\newblock}{\relax}
\providecommand{\bibinfo}[2]{#2}
\providecommand{\BIBentrySTDinterwordspacing}{\spaceskip=0pt\relax}
\providecommand{\BIBentryALTinterwordstretchfactor}{4}
\providecommand{\BIBentryALTinterwordspacing}{\spaceskip=\fontdimen2\font plus
\BIBentryALTinterwordstretchfactor\fontdimen3\font minus \fontdimen4\font\relax}
\providecommand{\BIBforeignlanguage}[2]{{%
\expandafter\ifx\csname l@#1\endcsname\relax
\typeout{** WARNING: IEEEtran.bst: No hyphenation pattern has been}%
\typeout{** loaded for the language `#1'. Using the pattern for}%
\typeout{** the default language instead.}%
\else
\language=\csname l@#1\endcsname
\fi
#2}}
\providecommand{\BIBdecl}{\relax}
\BIBdecl

\bibitem{vaswani2017attention}
A.~Vaswani, N.~Shazeer, N.~Parmar, J.~Uszkoreit, L.~Jones, A.~N. Gomez, {\L}.~Kaiser, and I.~Polosukhin, ``Attention is all you need,'' \emph{Advances in neural information processing systems}, vol.~30, 2017.

\bibitem{wolf2020transformers}
T.~Wolf, L.~Debut, V.~Sanh, J.~Chaumond, C.~Delangue, A.~Moi, P.~Cistac, T.~Rault, R.~Louf, M.~Funtowicz \emph{et~al.}, ``Transformers: State-of-the-art natural language processing,'' in \emph{Proceedings of the 2020 conference on empirical methods in natural language processing: system demonstrations}, 2020, pp. 38--45.

\bibitem{kalyan2021ammus}
K.~S. Kalyan, A.~Rajasekharan, and S.~Sangeetha, ``Ammus: A survey of transformer-based pretrained models in natural language processing,'' \emph{arXiv preprint arXiv:2108.05542}, 2021.

\bibitem{wolf2019huggingface}
T.~Wolf, L.~Debut, V.~Sanh, J.~Chaumond, C.~Delangue, A.~Moi, P.~Cistac, T.~Rault, R.~Louf, M.~Funtowicz \emph{et~al.}, ``Huggingface's transformers: State-of-the-art natural language processing,'' \emph{arXiv preprint arXiv:1910.03771}, 2019.

\bibitem{wu2020visual}
B.~Wu, C.~Xu, X.~Dai, A.~Wan, P.~Zhang, Z.~Yan, M.~Tomizuka, J.~Gonzalez, K.~Keutzer, and P.~Vajda, ``Visual transformers: Token-based image representation and processing for computer vision,'' \emph{arXiv preprint arXiv:2006.03677}, 2020.

\bibitem{bi2021transformer}
J.~Bi, Z.~Zhu, and Q.~Meng, ``Transformer in computer vision,'' in \emph{2021 IEEE International conference on computer science, electronic information engineering and intelligent control technology (CEI)}.\hskip 1em plus 0.5em minus 0.4em\relax IEEE, 2021, pp. 178--188.

\bibitem{liu2021swin}
Z.~Liu, Y.~Lin, Y.~Cao, H.~Hu, Y.~Wei, Z.~Zhang, S.~Lin, and B.~Guo, ``Swin transformer: Hierarchical vision transformer using shifted windows,'' in \emph{Proceedings of the IEEE/CVF international conference on computer vision}, 2021, pp. 10\,012--10\,022.

\bibitem{dong2018speech}
L.~Dong, S.~Xu, and B.~Xu, ``Speech-transformer: a no-recurrence sequence-to-sequence model for speech recognition,'' in \emph{2018 IEEE international conference on acoustics, speech and signal processing (ICASSP)}.\hskip 1em plus 0.5em minus 0.4em\relax IEEE, 2018, pp. 5884--5888.

\bibitem{gulati2020conformer}
A.~Gulati, J.~Qin, C.-C. Chiu, N.~Parmar, Y.~Zhang, J.~Yu, W.~Han, S.~Wang, Z.~Zhang, Y.~Wu \emph{et~al.}, ``Conformer: Convolution-augmented transformer for speech recognition,'' \emph{arXiv preprint arXiv:2005.08100}, 2020.

\bibitem{zhang2020transformer}
Q.~Zhang, H.~Lu, H.~Sak, A.~Tripathi, E.~McDermott, S.~Koo, and S.~Kumar, ``Transformer transducer: A streamable speech recognition model with transformer encoders and rnn-t loss,'' in \emph{ICASSP 2020-2020 IEEE International Conference on Acoustics, Speech and Signal Processing (ICASSP)}.\hskip 1em plus 0.5em minus 0.4em\relax IEEE, 2020, pp. 7829--7833.

\bibitem{powerai}
H.~M. L. C.~C. Power and D.~A.~I. Progress, ``Ai and compute.''

\bibitem{strubell2020energy}
E.~Strubell, A.~Ganesh, and A.~McCallum, ``Energy and policy considerations for modern deep learning research,'' in \emph{Proceedings of the AAAI conference on artificial intelligence}, vol.~34, no.~09, 2020, pp. 13\,693--13\,696.

\bibitem{patterson2021carbon}
D.~Patterson, J.~Gonzalez, Q.~Le, C.~Liang, L.-M. Munguia, D.~Rothchild, D.~So, M.~Texier, and J.~Dean, ``Carbon emissions and large neural network training,'' \emph{arXiv preprint arXiv:2104.10350}, 2021.

\bibitem{openai2023gpt4}
\BIBentryALTinterwordspacing
OpenAI, ``Gpt-4 technical report,'' 2023, accessed: 2023-10-31. [Online]. Available: \url{https://openai.com/research/gpt-4}
\BIBentrySTDinterwordspacing

\bibitem{volta}
NVIDIA-Tuning, ``Nvidia volta gpu architecture tuning guide.'' Oct. 2023, https://docs.nvidia.com/cuda/volta-tuning-guide/index.html.

\bibitem{dao2022flashattention}
T.~Dao, D.~Fu, S.~Ermon, A.~Rudra, and C.~R{\'e}, ``Flashattention: Fast and memory-efficient exact attention with io-awareness,'' \emph{Advances in Neural Information Processing Systems}, vol.~35, pp. 16\,344--16\,359, 2022.

\bibitem{tensor}
{Nvidia}, ``{ Tensor Core. },'' 2023, https://www.nvidia.cn/data-center/tensor-cores/.

\bibitem{dao2023flashattention}
T.~Dao, ``Flashattention-2: Faster attention with better parallelism and work partitioning,'' \emph{arXiv preprint arXiv:2307.08691}, 2023.

\bibitem{mma}
{NVIDIA}, ``{ Warp Level Matrix Multiply-Accumulate Instructions. },'' Aug. 2023, https://docs.nvidia.com/cuda/parallel-thread-execution/index.html\#matrix-shape.

\bibitem{milakov2018online}
M.~Milakov and N.~Gimelshein, ``Online normalizer calculation for softmax,'' \emph{arXiv preprint arXiv:1805.02867}, 2018.

\bibitem{rabe2021self}
M.~N. Rabe and C.~Staats, ``Self-attention does not need $o(n^2)$ memory,'' \emph{arXiv preprint arXiv:2112.05682}, 2021.

\bibitem{kitaev2020reformer}
N.~Kitaev, {\L}.~Kaiser, and A.~Levskaya, ``Reformer: The efficient transformer,'' \emph{arXiv preprint arXiv:2001.04451}, 2020.

\bibitem{pytorch}
{Pytorch}, ``{ Pytorch framework. },'' 2023, https://pytorch.org/docs/stable/index.html.

\bibitem{cublas}
\BIBentryALTinterwordspacing
N.~Corporation, \emph{cuBLAS Library User Guide}, 2023, version 12.0. [Online]. Available: \url{https://docs.nvidia.com/cuda/cublas/index.html}
\BIBentrySTDinterwordspacing

\bibitem{turing}
NVIDIA-Tuning, ``Nvidia turing gpu architecture tuning guide.'' Oct. 2023, https://docs.nvidia.com/cuda/turing-tuning-guide/index.html.

\bibitem{ampere}
------, ``Nvidia ampere gpu architecture tuning guide.'' Oct. 2023, https://docs.nvidia.com/cuda/ampere-tuning-guide/index.html.

\bibitem{ada}
NVIDIA-Ada, ``Nvidia ada gpu architecture tuning guide.'' Oct. 2023, https://docs.nvidia.com/cuda/ada-tuning-guide/index.html.

\bibitem{hopper}
NVIDIA-Hopper, ``Nvidia hopper tuning guide.'' Oct. 2023, https://docs.nvidia.com/cuda/hopper-tuning-guide/index.html.

\bibitem{ptx}
{Nvidia}, ``Ptx: Parallel thread execution,'' 2024, https://docs.nvidia.com/cuda/parallel-thread-execution/index.html.

\bibitem{devlin2018bert}
J.~Devlin, M.-W. Chang, K.~Lee, and K.~Toutanova, ``Bert: Pre-training of deep bidirectional transformers for language understanding,'' \emph{arXiv preprint arXiv:1810.04805}, 2018.

\bibitem{brown2020language}
T.~Brown, B.~Mann, N.~Ryder, M.~Subbiah, J.~D. Kaplan, P.~Dhariwal, A.~Neelakantan, P.~Shyam, G.~Sastry, A.~Askell \emph{et~al.}, ``Language models are few-shot learners,'' \emph{Advances in neural information processing systems}, vol.~33, pp. 1877--1901, 2020.

\bibitem{raffel2020exploring}
C.~Raffel, N.~Shazeer, A.~Roberts, K.~Lee, S.~Narang, M.~Matena, Y.~Zhou, W.~Li, and P.~J. Liu, ``Exploring the limits of transfer learning with a unified text-to-text transformer,'' \emph{Journal of machine learning research}, vol.~21, no. 140, pp. 1--67, 2020.

\bibitem{liu2019roberta}
Y.~Liu, M.~Ott, N.~Goyal, J.~Du, M.~Joshi, D.~Chen, O.~Levy, M.~Lewis, L.~Zettlemoyer, and V.~Stoyanov, ``Roberta: A robustly optimized bert pretraining approach,'' \emph{arXiv preprint arXiv:1907.11692}, 2019.

\bibitem{yang2019xlnet}
Z.~Yang, Z.~Dai, Y.~Yang, J.~Carbonell, R.~R. Salakhutdinov, and Q.~V. Le, ``Xlnet: Generalized autoregressive pretraining for language understanding,'' \emph{Advances in neural information processing systems}, vol.~32, 2019.

\bibitem{clark2020electra}
K.~Clark, M.-T. Luong, Q.~V. Le, and C.~D. Manning, ``Electra: Pre-training text encoders as discriminators rather than generators,'' \emph{arXiv preprint arXiv:2003.10555}, 2020.

\bibitem{radford2019language}
A.~Radford, J.~Wu, R.~Child, D.~Luan, D.~Amodei, I.~Sutskever \emph{et~al.}, ``Language models are unsupervised multitask learners,'' \emph{OpenAI blog}, vol.~1, no.~8, p.~9, 2019.

\bibitem{pybind11}
\BIBentryALTinterwordspacing
W.~Jakob, J.~Rhinelander, and D.~Moldovan, ``pybind11 – seamless operability between c++11 and python,'' 2017, accessed: 2024-10-31. [Online]. Available: \url{https://github.com/pybind/pybind11}
\BIBentrySTDinterwordspacing

\bibitem{beltagy2020longformer}
I.~Beltagy, M.~E. Peters, and A.~Cohan, ``Longformer: The long-document transformer,'' \emph{arXiv preprint arXiv:2004.05150}, 2020.

\bibitem{xFormers2022}
B.~Lefaudeux, F.~Massa, D.~Liskovich, W.~Xiong, V.~Caggiano, S.~Naren, M.~Xu, J.~Hu, M.~Tintore, S.~Zhang, P.~Labatut, D.~Haziza, L.~Wehrstedt, J.~Reizenstein, and G.~Sizov, ``xformers: A modular and hackable transformer modelling library,'' \url{https://github.com/facebookresearch/xformers}, 2022.

\bibitem{zaheer2020big}
M.~Zaheer, G.~Guruganesh, K.~A. Dubey, J.~Ainslie, C.~Alberti, S.~Ontanon, P.~Pham, A.~Ravula, Q.~Wang, L.~Yang \emph{et~al.}, ``Big bird: Transformers for longer sequences,'' \emph{Advances in neural information processing systems}, vol.~33, pp. 17\,283--17\,297, 2020.

\bibitem{wolf-etal-2020-transformers}
\BIBentryALTinterwordspacing
T.~Wolf, L.~Debut, V.~Sanh, J.~Chaumond, C.~Delangue, A.~Moi, P.~Cistac, T.~Rault, R.~Louf, M.~Funtowicz, J.~Davison, S.~Shleifer, P.~von Platen, C.~Ma, Y.~Jernite, J.~Plu, C.~Xu, T.~L. Scao, S.~Gugger, M.~Drame, Q.~Lhoest, and A.~M. Rush, ``Transformers: State-of-the-art natural language processing,'' in \emph{Proceedings of the 2020 Conference on Empirical Methods in Natural Language Processing: System Demonstrations}.\hskip 1em plus 0.5em minus 0.4em\relax Online: Association for Computational Linguistics, Oct. 2020, pp. 38--45. [Online]. Available: \url{https://www.aclweb.org/anthology/2020.emnlp-demos.6}
\BIBentrySTDinterwordspacing

\bibitem{ying2021lazyformer}
C.~Ying, G.~Ke, D.~He, and T.-Y. Liu, ``Lazyformer: Self attention with lazy update,'' \emph{arXiv preprint arXiv:2102.12702}, 2021.

\bibitem{fan2019reducing}
A.~Fan, E.~Grave, and A.~Joulin, ``Reducing transformer depth on demand with structured dropout,'' \emph{arXiv preprint arXiv:1909.11556}, 2019.

\bibitem{zhang2020accelerating}
M.~Zhang and Y.~He, ``Accelerating training of transformer-based language models with progressive layer dropping,'' \emph{Advances in neural information processing systems}, vol.~33, pp. 14\,011--14\,023, 2020.

\bibitem{choromanski2020rethinking}
K.~Choromanski, V.~Likhosherstov, D.~Dohan, X.~Song, A.~Gane, T.~Sarlos, P.~Hawkins, J.~Davis, A.~Mohiuddin, L.~Kaiser \emph{et~al.}, ``Rethinking attention with performers,'' \emph{arXiv preprint arXiv:2009.14794}, 2020.

\bibitem{liu2019variance}
L.~Liu, H.~Jiang, P.~He, W.~Chen, X.~Liu, J.~Gao, and J.~Han, ``On the variance of the adaptive learning rate and beyond,'' \emph{arXiv preprint arXiv:1908.03265}, 2019.

\bibitem{gong2019efficient}
L.~Gong, D.~He, Z.~Li, T.~Qin, L.~Wang, and T.~Liu, ``Efficient training of bert by progressively stacking,'' in \emph{International conference on machine learning}.\hskip 1em plus 0.5em minus 0.4em\relax PMLR, 2019, pp. 2337--2346.

\bibitem{li2020shallow}
B.~Li, Z.~Wang, H.~Liu, Y.~Jiang, Q.~Du, T.~Xiao, H.~Wang, and J.~Zhu, ``Shallow-to-deep training for neural machine translation,'' \emph{arXiv preprint arXiv:2010.03737}, 2020.

\bibitem{yao2021adahessian}
Z.~Yao, A.~Gholami, S.~Shen, M.~Mustafa, K.~Keutzer, and M.~Mahoney, ``Adahessian: An adaptive second order optimizer for machine learning,'' in \emph{proceedings of the AAAI conference on artificial intelligence}, vol.~35, no.~12, 2021, pp. 10\,665--10\,673.

\bibitem{zhang2020fixed}
X.~Zhang, S.~Liu, R.~Zhang, C.~Liu, D.~Huang, S.~Zhou, J.~Guo, Q.~Guo, Z.~Du, T.~Zhi \emph{et~al.}, ``Fixed-point back-propagation training,'' in \emph{Proceedings of the IEEE/CVF conference on computer vision and pattern recognition}, 2020, pp. 2330--2338.

\bibitem{sun2019hybrid}
X.~Sun, J.~Choi, C.-Y. Chen, N.~Wang, S.~Venkataramani, V.~V. Srinivasan, X.~Cui, W.~Zhang, and K.~Gopalakrishnan, ``Hybrid 8-bit floating point (hfp8) training and inference for deep neural networks,'' \emph{Advances in neural information processing systems}, vol.~32, 2019.

\bibitem{nvidia2020apex}
NVIDIA, ``Nvidia apex,'' \url{https://github.com/NVIDIA/apex}, 2020.

\bibitem{rasley2020deepspeed}
J.~Rasley, S.~Rajbhandari, O.~Ruwase, and Y.~He, ``Deepspeed: System optimizations enable training deep learning models with over 100 billion parameters,'' in \emph{Proceedings of the 26th ACM SIGKDD International Conference on Knowledge Discovery \& Data Mining}, 2020, pp. 3505--3506.

\bibitem{fang2021turbotransformers}
J.~Fang, Y.~Yu, C.~Zhao, and J.~Zhou, ``Turbotransformers: an efficient gpu serving system for transformer models,'' in \emph{Proceedings of the 26th ACM SIGPLAN Symposium on Principles and Practice of Parallel Programming}, 2021, pp. 389--402.

\bibitem{yu2022orca}
G.-I. Yu, J.~S. Jeong, G.-W. Kim, S.~Kim, and B.-G. Chun, ``Orca: A distributed serving system for $\{$Transformer-Based$\}$ generative models,'' in \emph{16th USENIX Symposium on Operating Systems Design and Implementation (OSDI 22)}, 2022, pp. 521--538.

\bibitem{kwon2023efficient}
W.~Kwon, Z.~Li, S.~Zhuang, Y.~Sheng, L.~Zheng, C.~H. Yu, J.~Gonzalez, H.~Zhang, and I.~Stoica, ``Efficient memory management for large language model serving with pagedattention,'' in \emph{Proceedings of the 29th Symposium on Operating Systems Principles}, 2023, pp. 611--626.

\bibitem{chen2018tvm}
T.~Chen, T.~Moreau, Z.~Jiang, L.~Zheng, E.~Yan, H.~Shen, M.~Cowan, L.~Wang, Y.~Hu, L.~Ceze \emph{et~al.}, ``$\{$TVM$\}$: An automated $\{$End-to-End$\}$ optimizing compiler for deep learning,'' in \emph{13th USENIX Symposium on Operating Systems Design and Implementation (OSDI 18)}, 2018, pp. 578--594.

\end{thebibliography}

\end{document}